\documentclass[12pt]{article}
\usepackage{amsfonts,amsbsy,bm,euscript,mathrsfs}
\usepackage{amssymb,stmaryrd,faktor}
\usepackage[tbtags]{amsmath}
\usepackage[title,titletoc]{appendix}
\usepackage[bookmarks=true,colorlinks=true,linkcolor=blue,citecolor=blue,urlcolor=blue,bookmarksnumbered]{hyperref}
\usepackage{graphics}
\usepackage{tikz}
\usetikzlibrary{matrix,arrows}

\newcommand{\mathsym}[1]{{}}
\makeatletter
\renewcommand\section{\@startsection {section}{1}{\z@}
{-3.5ex \@plus -1ex \@minus -.2ex}
{2.3ex \@plus.2ex}
{\normalfont\large\bfseries}}
\renewcommand\subsection{\@startsection{subsection}{2}{\z@}
{-3.25ex\@plus -1ex \@minus -.2ex}
{1.5ex \@plus.2ex}
{\normalfont\normalsize\bfseries}}
\renewcommand\subsubsection{\@startsection{subsubsection}{2}{\z@}
{-3.25ex\@plus -1ex \@minus -.2ex}
{1.5ex \@plus.2ex}
{\normalfont\small\bfseries}}
\makeatother

\def\id{\protect{{1 \kern-.28em {\rm l}}}}
\def\be{\begin{eqnarray}}
\def\ee{\end{eqnarray}}

\def\foot{\footnote}
\def\bi{\bibitem}

\def\ha{{1 \over 2}}
\def\td{\tilde}
\def\ci{\cite}

 \def \J {{\mathcal J}}

\def\L{\Lambda}

\def\a{\alpha}
\def\b{\beta}

\def\del{\partial}
\def\a{\alpha}

\def\g{\gamma}
\def\s{\sigma}

\def\ov{\over}

\def\b{\beta}
\def\l{\lambda}

\def\foot{\footnote}

\def\ci{\cite}

\def\l{\lambda}

\def\td{\tilde}

\def\bi{\bibitem}
\def\la{\label}
\def\l{\lambda}
\def\foot{\footnote}

\def\sql{{\sqrt \l}}
\def\adss{$AdS_5 \times S^5~$ }
\newcommand{\rf}[1]{(\ref{#1})}
\def\ov{\over}

\def\ha{{1\ov 2}}

\def\no{\nonumber}
\def\J{\mathcal{J}}
\def\del{\partial}

\def\J{{\cal J}}

\def\bi{\bibitem}
\def\la{\label}
\def\l{\lambda}
\def\foot{\footnote}

\def\sql{{\sqrt \l}}

\def\adss{$AdS_5 \times S^5$\ }

\def\ov{\over}

\def\varpi{{\rm w}}

\def\ve{\varepsilon}

\def\s{\sigma}

\def\L{{\cal L}}

 \def \sql {\sqrt{\lambda}}

\def\hh{{\rm h}}

\def\ifo{\iffalse}
\def\edd{\end{document}}
\def\ha{{{\textstyle{1 \ov2}}}}

\def\sql{{\sqrt{\l}}}

\def\ket{\rangle}

\def\sql{\sqrt{\l}}

\def\bea{\be}
\def\eea{\ee}
\def\eqref{\rf}

  \def \J {{\cal J}}\def \sql {{\sqrt\lambda}}

\def\text{ {\textstyle} }

\def\adst{$AdS_3 \times S^3 \times T^4$ }

\def\pp{\textrm{p}}
\def\eata{\eta}
\def\So{{\textrm S}_{_1}}
\def\Sio{{\textrm S}_{_2}}
\def\xpr{{x'}}
\def\spm{{_{\!\pm}}}
\def\smp{{_{\!\mp}}}
\def\ga{{\rm a}}
\def\fnsv{\vspace{3pt}}

\begin{document}

\overfullrule=0pt
\parskip=2pt
\parindent=12pt
\headheight=0in \headsep=0in \topmargin=0in \oddsidemargin=0in

\vspace{ -3cm}
\thispagestyle{empty}
\vspace{-1cm}

\rightline{ Imperial-TP-AT-2013-03}
\rightline{ HU-EP-13/20}

\begin{center}
\vspace{1cm}
{\Large\bf Massive
S-matrix of $AdS_3 \times S^3\times T^4$
superstring theory \\
\vspace{0.2cm}
with mixed 3-form flux
\vspace{0.2cm}
}
\vspace{1.5cm}

{B. Hoare$^{a,}$\footnote{ben.hoare@physik.hu-berlin.de} and A.A. Tseytlin$^{b,}$\footnote{Also at Lebedev Institute, Moscow. tseytlin@imperial.ac.uk }}\\

\vskip 0.6cm

{\em $^{a}$ Institut f\"ur Physik, Humboldt-Universit\"at zu Berlin, \\ Newtonstra\ss e 15, D-12489 Berlin, Germany}

\vskip 0.3cm

{\em
$^{b}$ The Blackett Laboratory, Imperial College,
London SW7 2AZ, U.K.
}

\vspace{.2cm}

\end{center}

\begin{abstract}
\noindent
The type IIB supergravity $AdS_3 \times S^3\times T^4$ background with mixed
RR and NSNS 3-form fluxes is a near-horizon limit of
a non-threshold bound state of D5-D1 and NS5-NS1 branes. The corresponding superstring world-sheet theory
is expected to be integrable, opening the possibility of computing its exact spectrum for any
values of the coefficient $q$ of the NSNS flux and the string tension.
In arXiv:1303.1447 we have found the tree-level S-matrix for the massive BMN excitations
in this theory, which turned out to have a simple dependence on $q$.
Here, by analyzing the constraints of symmetry and integrability, we propose an exact massive-sector dispersion relation
and the exact S-matrix for this world-sheet theory. The S-matrix generalizes its recent construction in the $q=0$ case
in arXiv:1303.5995.
\end{abstract}

\newpage

\setcounter{equation}{0}
\setcounter{footnote}{0}
\setcounter{section}{0}

\def \q {{q}}
\def \ga {{\rm a}}
\def \hq {{\hat q}}
\def \sig {q}
\def \cog {\hat q}
\renewcommand{\theequation}{1.\arabic{equation}}
\setcounter{equation}{0}
\section{Introduction}\label{sec0}

This paper is a sequel to \ci{htb} in which we initiated the study of S-matrix for elementary massive excitations of
superstring theory on $AdS_3 \times S^3 \times T^4$ with mixed RR + NSNS 3-form flux parametrized by $ q\in (0,1) $.
This model interpolates between the purely RR flux case ($q=0$) described by a supercoset GS superstring
and the purely NSNS flux case ($q=1$), which can be described by a supersymmetric WZW model.
Its classical integrability \ci{cz} is expected to extend to the full quantum level and thus, like in the
$AdS_5 \times S^5$ case \ci{rev}, should allow for an exact solution for the string spectrum for any value of the string tension $\hh$
and the parameter $q$.
This should shed light on the corresponding dual 2-d CFT which is currently not understood beyond
its supersymmetry-protected BPS sector.

Here we shall extend the tree-level ($\hh \to \infty$) S-matrix (section \ref{sec1}) found in \ci{htb} from the superstring action
to the exact in $\hh$ result using symmetry algebra considerations \ci{ht2,htb} and generalizing the
pure RR ($q=0$) result of \ci{bssst}. A key idea is that while the superstring symmetry group and the
symmetry algebra of the S-matrix do not depend on $q$, the representation of the latter on particle states does
(section \ref{sec3}).
This leads to a $q$-modified exact ``magnon'' dispersion relation discussed in section \ref{sec4}.
This exact dispersion can be found also by discretizing the spatial world-sheet direction in
the quadratic part of the light-cone gauge string action
with the 1-d lattice step being $\hh^{-1}$, i.e. the inverse of string tension.\foot{As in the $q=0$ case, this
suggests that the large tension limit corresponds to a continuum world sheet, while the small tension limit should correspond to a discrete
``spin chain'' theory.\fnsv}

In the $q=0$ case \ci{bssst} the exact S-matrix (written in terms of the Zhukovsky variables $x^\pm$) is completely fixed, up to
two phases, by its symmetry algebra and satisfies the Yang-Baxter equation without the need to use explicit form of the dispersion relation.
To generalize to the $\q \neq 0$ case we find a new set of Zhukovsky variables
$x^\pm_{{_\pm}}$ (the $\spm$ subscripts correspond to positively/negatively charged states)
that are consistent with the $q$-modified dispersion relation and
in terms of which the representation parameters and the exact S-matrix take the same form as in the $q=0$ case (section \ref{sec5}).
The S-matrix satisfies the Yang-Baxter equation and involves four phases that remain to be determined by additional considerations.
We discuss a conjecture for the phases in terms of their $q=0$ values and discuss explicitly the relation to the AFS phase
in the leading semiclassical string theory limit.
Section \ref{secconc} contains a discussion of some open problems.

\renewcommand{\theequation}{2.\arabic{equation}}
\setcounter{equation}{0}
\section{S-matrix in the BMN limit}\label{sec1}

The tree-level S-matrix for the massive BMN modes of the superstring theory on
$AdS_3 \times S^3 \times T^4$ with mixed 3-form flux was found in \cite{htb}.
The flux is parametrized by $0 \leq q \leq 1 $,
with $q = 0\ (\hq=1)$ corresponding to the pure RR case and
$q=1\ (\hq=0)$ -- to the pure NSNS case,
\be \la{1}
q^2 + \hq^2 =1 \ , \ \ \ \ \ \ \ \ \ \ \hq = \sqrt{ 1 - q^2} \ . \ee
The second parameter is the string tension $\hh$ related to the radius of $AdS_3$ or $S^3$:
\begin{equation}\label{ef}
\hh = \frac{\sqrt{\lambda}}{2\pi} = \frac{R^2}{2\pi\a'} \ .
\end{equation}
The quantized coefficient of the WZ term in the action is related to $\q$ and $\hh$ as
\be \la{ka} \q \sql = k \ , \qquad {\rm i.e.} \qquad
k = 2\pi\, \hh\, \sig \ . \ee

The near-BMN expansion of the $AdS_3 \times S^3 \times T^4$
superstring action describes $4+4$ (bosonic + fermionic) modes with mass $\cog=\sqrt{ 1 - q^2}$, and $4+4$ massless modes.
The corresponding
S-matrix for the massive modes can
be written as the graded tensor product of two copies of an S-matrix describing the scattering of $2+2$ massive modes.
Let us denote the two massive bosons associated to the orthogonal directions of $S^3$
as the complex scalar $y= y_1 + iy_2$ and the corresponding scalar
for $AdS_3$ as $z=z_1 + iz_2$. The four massive fermions will be represented
as two complex Grassmann fields $\zeta$ and $\chi$.
Then we can define the following tensor product states
\begin{equation}\begin{split}\label{bhtens}
\left|y\right> = \left|\phi\right> \otimes \left|\phi\right> \ , \qquad & \left|z\right> = \left|\psi\right> \otimes \left|\psi\right> \ ,
\\
\left|\zeta\right> = \left|\phi\right> \otimes \left|\psi\right> \ , \qquad & \left|\chi\right> = \left|\psi\right> \otimes \left|\phi\right> \ ,
\end{split}\end{equation}
where $\left|\phi\right>$ is bosonic and $\left|\psi\right>$ is fermionic. The factorization property
means that the S-matrix for $\{y,z,\zeta,\chi\}$ can be constructed from an S-matrix for $\{\phi,\psi\}$,
which takes the following form
\begin{align} \nonumber
\mathbb{S}\left|\phi_\pm\phi_\pm{}'\right> = & A_\spm L_{1_\pm} \left|\phi_\pm\phi_\pm{}'\right>\ ,
& \mathbb{S}\left|\phi_\pm\psi_\pm{}'\right> = & A_\spm L_{3_\pm}\left|\phi_\pm\psi_\pm{}'\right> + A_\spm L_{5_\pm}\left|\psi_\pm\phi_\pm{}'\right>\ ,
\\ \nonumber
\mathbb{S}\left|\psi_\pm\psi_\pm{}'\right> = & A_\spm \Lambda_{1_\pm}\left|\psi_\pm\psi_\pm{}'\right>\ ,
& \mathbb{S}\left|\psi_\pm\phi_\pm{}'\right> = & A_\spm\Lambda_{3_\pm}\left|\psi_\pm\phi_\pm{}'\right> + A_\spm \Lambda_{5_\pm} \left|\phi_\pm\psi_\pm{}'\right>\ ,
\\ \nonumber
\mathbb{S}\left|\phi_\pm\psi_\mp{}'\right> = &\bar A_\spm L_{6_\pm} \left|\phi_\pm\psi_\mp{}'\right>\ ,
& \mathbb{S}\left|\phi_\pm\phi_\mp{}'\right> = &\bar A_\spm L_{2_\pm}\left|\phi_\pm\phi_\mp{}'\right> + \bar A_\spm L_{4_\pm}\left|\psi_\pm\psi_\mp{}'\right>\ ,
\\ \label{bct}
\mathbb{S}\left|\psi_\pm\phi_\mp{}'\right> = & \bar A_\spm \Lambda_{6_\pm}\left|\psi_\pm\phi_\mp{}'\right>\ ,
& \mathbb{S}\left|\psi_\pm\psi_\mp{}'\right> = & \bar A_\spm \Lambda_{2_\pm}\left|\psi_\pm\psi_\mp{}'\right> + \bar A_\spm \Lambda_{4_\pm} \left|\phi_\pm\phi_\mp{}'\right>\ ,
\end{align}
where the signs $\pm$ represent the charges, i.e. correspond to the fields and their conjugates.
\foot{Explicitly, we shall use the following notation: $\phi_+ =\phi, \ \phi_- =\phi^*; \ \psi_+ =\psi, \ \phi_- =\psi^*$.
In \cite{bor1,bssst} the ``$+$''-sector was referred to as ``left'' (L) and the
``$-$''-sector as ``right'' (R). We shall not follow this terminology here as it is somewhat confusing:
the LL, LR, etc., notation for S-matrices is usually reserved for the massless scattering case.\label{foot2}\fnsv}
The structure of this S-matrix
\rf{bct} is constrained by the requirement of a $U(1)^2$ symmetry under which $\{\phi,\psi\}$ have charges $\{1,0\}$ and $\{0,1\}$ respectively.
The leading-order term in the expansion in the inverse string tension $\hh^{-1}$ gives the tree-level S-matrix, for which
the phases
\begin{equation}
A_\spm = 1-\frac i{2\hh}\big(\ga-\tfrac12\big)(e_\pm'p-e_\pm p') + \mathcal{O}(\hh^{-2})\ , \qquad
\bar A_\spm = 1-\frac i{2\hh}\big(\ga-\tfrac12\big)(e_\mp'p-e_\pm p') + \mathcal{O}(\hh^{-2})\ \la{pha}
\end{equation}
contain the
dependence on the gauge parameter $\ga$
($\ga=\ha$ corresponds to the uniform light-cone gauge)\foot{Here we use the notation $\ga$ for the gauge parameter instead of $a$ used in
\ci{htb} and some earlier references.\fnsv}
and the other non-trivial functions of the momenta $p, p'$ and energies $e_\pm,e'_\pm$ are given by \ci{htb}
\begin{align}
& L_{1_\pm} = 1 + \frac{i}{2\hh}l_{1_{\pm}} + \mathcal{O}(\hh^{-2}) \ , && \Lambda_{1_\pm} = 1 - \frac{i}{2\hh} l_{1_{\pm}} + \mathcal{O}(\hh^{-2}) \ , && \nonumber
\\
& L_{3_\pm} = 1 + \frac{i}{2\hh}l_{3_{\pm}} + \mathcal{O}(\hh^{-2}) \ , && \Lambda_{3_\pm} = 1 - \frac{i}{2\hh} l_{3_{\pm}} + \mathcal{O}(\hh^{-2}) \ , && \nonumber
\\
& L_{6_\pm} = 1 + \frac{i}{2\hh}l_{3_{\pm}} + \mathcal{O}(\hh^{-2}) \ , && \Lambda_{6_\pm} = 1 - \frac{i}{2\hh} l_{3_{\pm}} + \mathcal{O}(\hh^{-2}) \ , && \nonumber
\\
& L_{2_\pm} = 1 + \frac{i}{2\hh}l_{2_{\pm}} + \mathcal{O}(\hh^{-2}) \ , && \Lambda_{2_\pm} = 1 - \frac{i}{2\hh} l_{2_{\pm}} + \mathcal{O}(\hh^{-2}) \ , && \nonumber
\\
& L_{5_\pm} = - \frac{i}{\hh} l_{5_{\pm}} + \mathcal{O}(\hh^{-2}) \ , && \Lambda_{5_\pm} = -\frac{i}{\hh} l_{5_{\pm}} + \mathcal{O}(\hh^{-2}) \ ,\la{fu} &&
\\
& L_{4_\pm} = \frac{i}{\hh} l_{4_{\pm}} + \mathcal{O}(\hh^{-2}) \ , && \Lambda_{4_\pm} = \frac{i}{\hh} l_{4_{\pm}} + \mathcal{O}(\hh^{-2}) \ , && \nonumber
\end{align}
where
\begin{align}
& l_{1_{\pm}} = \frac{(p+p')(e_\pm'p+e_\pm p')}{2(p-p')} \ ,
\qquad \qquad
l_{2_{\pm}} = \frac{(p-p')(e_\mp'p+e_\pm p')}{2(p+p')}\ , \nonumber
\\
&l_{3_{\pm}} = - \frac{1}{2}(e_\pm'p+e_\pm p') \ , \nonumber
\\
& l_{4_{\pm}} = - \frac{pp'}{2(p+p')}\Big[\sqrt{(e_\pm+p\pm \sig)(e_\mp'+p'\mp \sig)} - \sqrt{(e_\pm-p\mp \sig)(e_\mp'- p'\pm \sig)}\, \Big] \ , \nonumber
\\
& l_{5_{\pm}} = - \frac{pp'}{2(p-p')}\Big[\sqrt{(e_\pm+p\pm \sig)(e_\pm'+p'\pm \sig)} + \sqrt{(e_\pm-p\mp \sig)(e_\pm'-p'\mp \sig )}\, \Big] \ , \la{lel}
\\
& e_\pm = \sqrt{\hq^2
+ (p\pm \sig)^2 } \ , \qquad \qquad \qquad \ \,
e'_\pm = \sqrt{\hq^2
+ (p'\pm \sig){}^2} \label{1_5} \ .
\end{align}
Eq.\rf{1_5} gives the dispersion relation (which is the same for the bosonic and fermionic modes),
$ e^2_\pm = 1 + p^2 \pm 2 \sig \, p $, \ generalizing the familiar BMN massive relativistic dispersion relation.
The energy is minimized when $p=\mp \sig$ so that $ \cog $ is the mass of the corresponding excitations.

For the case of pure RR flux ($\q = 0$)
an all-loop
conjecture for the S-matrix for the massive modes was made in \cite{bssst} using symmetry algebra considerations
and
integrability constraints.\foot{The phase factors still remain to be explicitly determined
from crossing condition and the requirement of correspondence with string perturbation theory, see a discussion in \ci{bssst}.\fnsv}
The aim of the present work is to
extend the tree-level $q\not=0$ result \ci{htb} for the S-matrix to all orders in $\hh^{-1}$,
i.e. to generalize the exact S-matrix proposal of \ci{bssst} to
the presence of a non-vanishing NSNS flux.
The starting point will be to understand how the action of the symmetry algebra of the S-matrix
is modified for $\q\not=0$.

\renewcommand{\theequation}{3.\arabic{equation}}
\setcounter{equation}{0}
\section{The S-matrix symmetry algebra and its representation}\label{sec3}

The type IIB supergravity background corresponding to the superstring under consideration
is the near-horizon limit of the non-threshold BPS bound state of NS5-NS1 and D5-D1 branes
(see, e.g., \ci{ruts} and references there) and can thus be obtained, e.g., by applying S-duality
to the NS5-NS1 ($\q= 1$)
or D5-D1 ($\q = 0$)
solution. This means that the space-time symmetry
of this background can not depend on $\q$.
Indeed, the non-trivial ``massive'' $AdS_3\times S^3$ part of the superstring action can be described
by the same supercoset geometry $[PSU(1,1|2) \times PSU(1,1|2)]/[SU(1,1) \times SU(2)]$ \ci{pes}
with $\q$
appearing only as a parameter in the action \ci{cz}.

For this reason it is not surprising that the symmetry algebra of the corresponding S-matrix
(which should be a subalgebra of the supercoset symmetry preserved by the BMN vacuum)
should not depend on $\q$.
However, the dependence on $\q$
may enter the form of its representation on particle states.

For $\q = 0$
the factorized form of the S-matrix described in section \ref{sec1} is a consequence of the structure of the
symmetry algebra and the integrability. As the theory should be integrable for any $\q$,
the exact S-matrix should also factorize. Furthermore, the factor S-matrix should satisfy the Yang-Baxter equation.

The symmetry algebra here is also the same as in the case of the S-matrix
of the Pohlmeyer-reduced theory corresponding to the $AdS_3 \times S^3$ superstring \ci{ht2,gt}.
Its
generators are: $(i)$ two $U(1)$ generators $\mathfrak R$ and $\mathfrak L$; $(ii)$ four supercharges
$\mathfrak Q_{\pm\mp}$ and $\mathfrak S_{\pm\mp}$ ($+$ and $-$ denote the
charges under the $U(1) \times U(1)$ bosonic subalgebra); $(iii)$
three central extension generators
$\mathfrak C$, $\mathfrak P$ and $\mathfrak K$. Defining
\begin{equation}
\mathfrak{M} =\frac12 (\mathfrak R + \mathfrak L) \ , \qquad \mathfrak B =\frac12( \mathfrak R - \mathfrak L)\ ,
\end{equation}
the non-vanishing (anti-)commutation relations are given by
\begin{align}
&[ \mathfrak B, \, \mathfrak Q_{\pm\mp} ] = \pm i \mathfrak Q_{\pm\mp} \ , & \nonumber & [ \mathfrak B, \, \mathfrak S_{\pm\mp} ] = \pm i \mathfrak S_{\pm\mp} \ , &&
\\
& \{\mathfrak Q_{\pm\mp} , \,\mathfrak Q_{\mp\pm}\} = \mathfrak P \ , & \label{bcd} & \{\mathfrak S_{\pm\mp} , \,\mathfrak S_{\mp\pm}\} = \mathfrak K \ , &&
\{\mathfrak Q_{\pm\mp} , \,\mathfrak S_{\mp\pm}\} =
\pm i\, \mathfrak M + \mathfrak C \ .
\end{align}
These are consistent with the following set of reality conditions
\begin{equation}\label{bhrc}
\mathfrak B^\dagger = -\mathfrak B \ , \qquad \mathfrak Q_{\pm\mp}^\dagger = \mathfrak S_{\mp\pm} \ , \qquad \mathfrak M^\dagger = -\mathfrak M \ ,\qquad \mathfrak P^\dagger = \mathfrak K \ , \qquad \mathfrak C^\dagger = \mathfrak C \ .
\end{equation}
This superalgebra is a centrally-extended semi-direct sum of $\mathfrak u(1)$ (generated by $\mathfrak B$)
with two copies of the superalgebra $\mathfrak{psu}(1|1)$, i.e.
\begin{equation}
[\mathfrak u(1)\inplus \mathfrak{psu}(1|1)^2 ]\ltimes \mathfrak u(1) \ltimes \mathbb{R}^3 \ . \la{al}
\end{equation}
The central extensions are represented by the generators $\mathfrak M$, $\mathfrak C$, $\mathfrak P$ and $\mathfrak K$.
As for the other three central extensions, $\mathfrak C$, $\mathfrak P$ and $\mathfrak K$, there is
therefore only a single copy of the $\mathfrak u(1)$ central extension $\mathfrak M$ when we consider the symmetry of the full S-matrix
\begin{equation}
[\mathfrak u(1)\inplus \mathfrak{psu}(1|1)^2]^2\ltimes \mathfrak u(1) \ltimes \mathbb{R}^3 \ . \la{als}
\end{equation}

The algebra \eqref{bcd} is a subalgebra of the familiar $\mathfrak{psu}(2|2)\ltimes \mathbb{R}^3$ which was
a factor symmetry in the
corresponding construction of the S-matrix of the $AdS_5 \times S^5$ superstring theory \ci{beis1,afr}.
In that case the automorphism group of the algebra was $SL(2,\mathbb{C})$; here, however,
it is enhanced to $GL(2,\mathbb{C})$. The additional $GL(1,\mathbb{C})$
acts as follows:
\begin{equation}
(\mathfrak B, \ \mathfrak Q_{\pm\mp}, \ \mathfrak S_{\pm\mp}, \ \mathfrak M, \ \mathfrak C,\ \mathfrak P, \ \mathfrak K)
\rightarrow \
(\mathfrak B , \ \nu \, \mathfrak Q_{\pm\mp}, \ \nu \, \mathfrak S_{\pm\mp}, \ \nu^2 \, \mathfrak M, \ \nu^2 \, \mathfrak C, \ \nu^2 \, \mathfrak P, \ \nu^2\, \mathfrak K) \label{automorphism}
\ . \end{equation}
The presence of this additional automorphism, in principle, allows one
to introduce
different mass values
(as is the case, e.g., for the $AdS_3 \times S^3 \times S^3 \times S^1$ theory with pure RR flux \ci{bsz})
through rescaling the eigenvalue of $\mathfrak M$.

\

The particular representation of this symmetry algebra which is
of interest to us here
consists of one complex boson $\phi$ and one complex fermion $\psi$.
The generators have the following action on the one-particle states
(we use the notation $\phi_+ =\phi, \ \phi_- =\phi^*; \ \psi_+ =\psi, \ \phi_- =\psi^*$)
\begin{align}
& \mathfrak B\left|\phi_\pm\right>=\pm i\left|\phi_\pm\right> \ , & & \nonumber & & \mathfrak B\left|\psi_\pm\right>=\mp i \left|\psi_\pm\right> \ ,
\\
& \mathfrak Q_{\pm\mp}\left|\phi_\pm\right>=0 \ , & & \nonumber & & \mathfrak Q_{\pm\mp}\left|\psi_\pm\right>=b_\pm \left|\phi_\pm\right> \ ,
\\
& \mathfrak Q_{\mp\pm}\left|\phi_\pm\right>=a_\pm\left|\psi_\pm\right> \ , & & \nonumber & & \mathfrak Q_{\mp\pm}\left|\psi_\pm\right>=0 \ ,
\\
& \mathfrak S_{\pm\mp}\left|\phi_\pm\right>=0 \ , & & \nonumber & & \mathfrak S_{\pm\mp}\left|\psi_\pm\right>=d_\pm\left|\phi_\pm\right> \ ,
\\
& \mathfrak S_{\mp\pm}\left|\phi_\pm\right>=c_\pm \left|\psi_\pm\right> \ , & & \nonumber & & \mathfrak S_{\mp\pm}\left|\psi_\pm\right>=0 \ ,
\\
& \mathfrak M\left|\phi_\pm\right>=\pm \tfrac i2 M_\pm \left|\phi_\pm \right> \ , & & \nonumber & & \mathfrak M\left|\psi_\pm\right>= \pm \tfrac i2 M_\pm \left|\psi_\pm\right> \ ,
\\
& \mathfrak C\left|\phi_\pm\right>= C_\pm \left|\phi_\pm \right> \ , & & \nonumber & & \mathfrak C\left|\psi_\pm\right>= C_\pm \left|\psi_\pm\right> \ ,
\\
& \mathfrak P\left|\phi_\pm\right>=P_\pm \left|\phi_\pm \right> \ , & & \nonumber & & \mathfrak P\left|\psi_\pm\right>= P_\pm \left|\psi_\pm\right> \ ,
\\
& \mathfrak K\left|\phi_\pm\right>= K_\pm \left|\phi_\pm \right> \ , & & & & \mathfrak K\left|\psi_\pm\right>= K_\pm \left|\psi_\pm\right> \ .
\label{bhabcd}
\end{align}
Here $a_\pm,b_\pm,c_\pm,d_\pm,C_\pm,P_\pm$ and $K_\pm$ are the representation parameters
that will eventually be functions of the energy and
momentum of the state.
The fact that this representation is actually reducible ($\{\phi_+,\psi_+\}$ and $\{\phi_-,\psi_-\}$
are two irreducible representations related by conjugation --
called ``left'' and ``right'' in a related context \cite{bor1},
cf. footnote \ref{foot2})
allows for the introduction of the subscripts $\pm$ on the representation parameters.
The parameters with subscript ``$+$'' should
be related to those with subscript ``$-$'' by charge conjugation (see below).
The motivation for introducing the subscripts is clear from the tree-level
results of \cite{htb}, summarized above in section \ref{sec1}.

For the supersymmetry algebra to close the following conditions should be satisfied
\begin{equation}\begin{split}\label{bhabcdpkc}
a_\pm b_\pm = P_\pm \ , \qquad c_\pm d_\pm = K_\pm \ , \qquad & a_\pm d_\pm = C_\pm + \frac{M_\pm}2 \ , \qquad b_\pm c_\pm = C_\pm-\frac{M_\pm}2\ .
\end{split}\end{equation}
These can easily be seen to imply that
\begin{equation}\label{bhshort}
C_\pm^2 = \frac{M_\pm^2}4 + P_\pm K_\pm \ ,
\end{equation}
which are just the shortening conditions for the two irreducible atypical representations. 
Physically, they will be interpreted as dispersion relations, with $M_\pm, C_\pm, P_\pm$ and $K_\pm$ defined in terms of the energy and momentum. In particular, in the near-BMN limit they should reduce to the expressions \eqref{1_5} found at leading order in perturbation theory.
The representation parameters are further constrained by the reality conditions \eqref{bhrc}
\begin{equation}\label{bhrcc}
a_\pm^* = d_\pm \ , \qquad b_\pm^* = c_\pm \ , \qquad M_\pm^* = M_\pm \ , \qquad C_\pm^* = C_\pm \ , \qquad P_\pm^* = K_\pm \ .
\end{equation}

To define the action of this symmetry on the two-particle states we need to introduce the coproduct
\begin{align}
\Delta(\mathfrak B) & = \mathfrak B \otimes \mathbb{I} + \mathbb{I} \otimes \mathfrak B \ , & \Delta(\mathfrak M) = & \mathfrak M \otimes \mathbb{I} + \mathbb{I} \otimes \mathfrak M \ , & \Delta(\mathfrak C) & = \mathfrak C \otimes \mathbb{I} + \mathbb{I} \otimes \mathfrak C \ , \nonumber
\\
\Delta(\mathfrak Q) & = \mathfrak Q \otimes \mathbb{I} + \mathfrak U \otimes \mathfrak Q \ , & & \nonumber & \Delta(\mathfrak S) & = \mathfrak S \otimes \mathbb{I} + \mathfrak U^{-1} \otimes \mathfrak S \ ,
\\
\Delta(\mathfrak P) & = \mathfrak P \otimes \mathbb{I} + \mathfrak{U}^2 \otimes \mathfrak P \ , & & & \Delta(\mathfrak K) & = \mathfrak K \otimes \mathbb{I} + \mathfrak{U}^{-2} \otimes \mathfrak K \ ,\label{coproduct}
\end{align}
and the opposite coproduct, defined as
\begin{equation}
\Delta^{\text{op}}(\mathfrak J) = \mathcal{P}(\Delta(\mathfrak J)) \ ,
\end{equation}
where $\mathfrak J$ is an arbitrary generator and $\mathcal{P}$ defines the graded permutation of the tensor product.

The coproduct differs from the usual product by the introduction of a
new abelian generator $\mathfrak U$,
with $\Delta(\mathfrak U) = \mathfrak U \otimes \mathfrak U$ \cite{tor}
(see also \ci{ht2}).
This is done according to a $\mathbb{Z}$-grading of the algebra, whereby the charges $-2,-1,1,2$ are associated to the generators $\mathfrak K$, $\mathfrak S$, $\mathfrak Q$, $\mathfrak P$
while the remaining generators are uncharged. The action of $\mathfrak U$ on the single-particle states is given by
\begin{equation}
\mathfrak U\left|\phi_\pm\right> = U_\pm \left|\phi_\pm\right> \ , \qquad\qquad
\mathfrak U\left|\psi_\pm\right> = U_\pm\left|\psi_\pm\right> \ .\la{uu}
\end{equation}
This braiding allows for the existence of a non-trivial S-matrix. It should be noted that for the central extensions the
coproduct should be equal to its opposite --- an issue we will return to in sections \ref{sec21} and \ref{sec22}.

The factorized tree-level S-matrix of the theory with mixed 3-form flux ($\q \neq 0$)
given in \rf{bct}--\rf{1_5}
co-commutes ($\Delta^{op}(\mathfrak J) \, \mathbb S = \mathbb S \, \Delta(\mathfrak J)$) with the supersymmetry algebra if the
representation parameters in \rf{bhabcd},\rf{uu} have the following
form at the leading order
in the large $\hh$ (near-BMN)
expansion
\begin{align}
a_\pm =\, &\frac{e^{-\frac{i\pi}4}}{\sqrt2}\sqrt{e_\pm + 1 \pm \sig \,p} \ , &
b_\pm =\, &-\frac{ie^{\frac{i\pi}4}}{\sqrt2} \frac{\hq \,p}{\sqrt{e_\pm +1 \pm \sig\,p}} \ , \nonumber
\\
c_\pm =\, &\frac{ie^{-\frac{i\pi}4}}{\sqrt2} \frac{\hq \,p}{\sqrt{e_\pm +1 \pm \sig\,p}} \ , &
d_\pm =\, &\frac{e^{\frac{i\pi}4}}{\sqrt2}\sqrt{e_\pm + 1 \pm \sig \,p} \ ,\nonumber
\\
U_\pm = \, & \, 1 + \frac{ip}{2\hh} \ , \quad \ M_\pm = 1\pm \sig \, p \ , \quad \ C_\pm =\, \frac{e_\pm}{2} \ ,
& P_\pm =\, & - \frac{i}{2} \, \cog \, p \ , \quad \ K_\pm = \frac{i}{2}\, \cog \, p \ . \label{Functions}
\end{align}
$C_\pm$ thus plays the r\^ole of the energy.
In the
$q \to 0$
limit
$P_\pm$ and $K_\pm$ are proportional to the spatial momentum and $M_\pm$ is the effective mass
parameter, while
in the
$q \to 1$ limit $P_\pm$ and $K_\pm$ vanish, while
$M_\pm$ is the spatial momentum shifted by $\pm 1$.

\subsection{Exact expressions in the $\q=0$ case}\label{sec21}

To generalize the above expressions for the representation parameters
\rf{Functions} to all orders in $\hh$
let us first
review their algebraic construction for the pure RR case of $\q = 0$
\cite{htb,bssst}.
In this case the parameters with $+$ and $-$ subscripts are equal --- there
is a formal symmetry under the interchange of the two irreducible atypical representations;
therefore,
we will drop them for the remainder of this subsection.
The set of equations \eqref{bhabcdpkc} can be solved for $a,b,c,d$ in terms of $M,C,P$ and $K$ as
\begin{align}\nonumber
a =\, & \frac{\a \,e^{-\frac{i\pi}4}}{\sqrt2} \sqrt{2C + M} \ , &\
b =\, & \sqrt2\,\a^{-1} e^{\frac{i\pi}4} \frac{P}{\sqrt{2C + M}} \ , \\
c =\, & \sqrt2\, \a\, e^{-\frac{i\pi}4} \frac{K}{\sqrt{2C + M}} \ , &
d =\, & \frac{\a^{-1}e^{\frac{i\pi}4}}{\sqrt 2} \sqrt{2C +M} \ . \label{ad}
\end{align}
Here $\a$ is a phase parametrizing the normalization of the fermionic states with respect to the bosonic states,
and can be a function of the central extensions. To match
the expressions in \rf{Functions} corresponding to the tree-level string world-sheet
S-matrix \cite{htb} summarized in section \ref{sec1}
we should take $\a = 1 + \mathcal{O}(\hh^{-1})$.
To facilitate comparison with the literature, for the moment we will leave $\a$ unfixed.
Furthermore, we can use the $GL(1,\mathbb{C})$ automorphism \eqref{automorphism} to fix
\begin{equation}
M = 1\ . \la{mm}
\end{equation}
As was mentioned above, one important consequence of the non-trivial braiding
\eqref{coproduct} is that for the central extensions the coproduct should be
equal to its opposite. This implies
\begin{equation}
\mathfrak P \propto (1 - \mathfrak U^2) \ ,\qquad \qquad \mathfrak K \propto (1 - \mathfrak U^{-2}) \ . \la{pp}
\end{equation}
We fix the normalization of $\mathfrak P$ relative to $\mathfrak K$ by taking both constants of proportionality to be
equal to $\ha \hh$
where
the reality conditions \eqref{bhrcc} require that $\hh$ is real
($\hh$ will later be interpreted as
string tension).\footnote{The reality conditions \eqref{bhrcc} do allow for the introduction of an additional
phase into the constants of proportionality, i.e. $\ha \hh e^{i\varphi}$ and $\ha \hh e^{-i\varphi}$. However,
this phase does not appear in
the S-matrix and thus we set it equal to one.\fnsv}
Acting on the single-particle states gives us the relations\,\footnote{Note that the mapping between the representation parameters here and those used in
\cite{bssst}
is as follows:
\begin{equation*}
\{P,K,a,b,c,d\}_{here} = \{U^2 P,U^{-2} P^*,U a,U c,U^{-1} d,U^{-1} b\}_{there}\ , \label{footcompare1}
\end{equation*}
where we have denoted the eigenvalues of the generators $\mathfrak{P},\mathfrak{P}^\dagger$ in \cite{bssst} as $P,P^*$.\fnsv}
\begin{equation}\label{bhpku}
P = \frac \hh2 \, (1 - U^2) \ ,\qquad \qquad K = \frac \hh2 \, (1 - U^{-2}) \ ,
\end{equation}
where $U$ should satisfy, as a consequence of \eqref{bhrcc}, the following reality condition
\begin{equation}\label{bhurccp}
U^* = U^{-1} \ .
\end{equation}
Motivated by the well-known construction in the $AdS_5 \times S^5$ case
(implying a similar one in the \adst case with $\q=0$ \cite{bor1,bssst,htb})
we identify $C$ with (half) the energy and define $U$ in terms of the spatial momentum $\pp$ as
\begin{equation}\label{bhcuep}
C = \frac{e}2 \ ,\qquad \qquad U = e^{\frac i2 \pp} \ .
\end{equation}
Using \eqref{bhpku} and \eqref{bhcuep} we can substitute in for $C$, $P$ and $K$ in terms of the energy and the
momentum in the shortening conditions \eqref{bhshort} to find the following familiar dispersion relation
\begin{equation}\la{apiii}
e^2 = 1 + 4 \, \hh^2 \, \sin^2\frac \pp2 \ .
\end{equation}
Therefore, $\hh$ should be taken to be equal to the string tension eq.\eqref{ef}.
In terms of the energy and the momentum the representation parameters $a,b,c$ and $d$ \eqref{ad} are then given by
\begin{align}\nonumber
a =\, & \frac{\a \,e^{-\frac{i\pi}4}}{\sqrt2} \sqrt{e+1} \ , &\
b =\, & \frac{\a^{-1} e^{\frac{i\pi}4}}{\sqrt2} \frac{\hh(1-e^{i\pp})}{\sqrt{e+1}} \ , \\
c =\, & \frac{\a\, e^{-\frac{i\pi}4}}{\sqrt2} \frac{\hh(1-e^{-i\pp})}{\sqrt{e+1}} \ , &
d =\, & \frac{\a^{-1}e^{\frac{i\pi}4}}{\sqrt 2} \sqrt{e+1} \ . \label{bep}
\end{align}
We recall that $\a$ is a phase, which may depend on the momentum.

To define the near-BMN expansion we should set
\begin{equation}
\pp = \hh^{-1} p \ , \la{ppp}
\end{equation}
and expand the representation parameters
in powers of $\hh^{-1}$
keeping $p$ fixed.
Then the dispersion relation \rf{apiii} reduces to the BMN one, $e^2 =1 + p^2 $, and
setting $\a = 1 + \mathcal{O}(\hh^{-1})$
we find the agreement with the $\q \to 0$
limit of the expressions in \eqref{Functions}.
This means that
the factorized tree-level S-matrix of the theory with pure
RR flux ($\q = 0$)
co-commutes with the above symmetry algebra.

\subsection{Exact expressions in the \texorpdfstring{$\q\neq 0$}{q!=0} case}\label{sec22}

Let us
now generalize the above
algebraic construction to $\q \neq 0$.
Starting with the expressions \eqref{Functions} found
in the limit
\be \la{bmn}
\hh \to
\infty\ , \qquad \pp \to 0 \ , \ \qquad \qquad \ p\equiv \hh\, \pp= {\rm fixed } \ , \ee
it is natural to conjecture that the exact expressions for
$U_\pm,M_\pm,C_\pm, P_\pm$ and $K_\pm$
should be
\begin{equation} \begin{split}\label{uq}
& U_\pm = e^{\frac i2 \pp} \ , \qquad \qquad M_\pm = 1 \pm 2\hh\, \sig\, \sin \frac \pp2 \ ,\qquad \qquad C_\pm = \frac {e_\pm}{2} \ ,
\\ & P_\pm = \frac{\hh\,\cog}{2} \, (1 - e^{i \pp} ) \ ,\qquad \qquad \qquad \qquad \ \ \ \, K_\pm = \frac{\hh \,\cog}{2} \, (1 - e^{- i \pp} ) \ .
\end{split}\end{equation}
Then the algebraic requirement that the coproduct for $\mathfrak{P}$ and $\mathfrak{K}$ should be equal to its opposite is still satisfied.\foot{It should be noted that eq.\rf{uq} is not the most general solution of this requirement but is a simple one that seems physically motivated: it matches the ``discretization'' interpretation of the resulting dispersion relation discussed in section \ref{sec4}, suggestive of an underlying spin chain picture, by analogy with the $AdS_5 \times S^5$ case.\fnsv}
The corresponding expressions for $a_\pm,b_\pm,c_\pm$ and $d_\pm$ are therefore given by (cf. \rf{Functions})
\begin{align}\nonumber
a_\pm =\, & \frac{\a_\pm \,e^{-\frac{i\pi}4}}{\sqrt2} \sqrt{e_\pm+1\pm2\hh\,\q\,\sin\tfrac\pp2} \ , &\
\ \ \ b_\pm =\, & \frac{\a^{-1}_\pm e^{\frac{i\pi}4}}{\sqrt2} \frac{\hh\,\hq\,(1-e^{i\pp})}{\sqrt{e_\pm+1\pm2\hh\,\q\,\sin\frac\pp2}} \ , \\
c_\pm =\, & \frac{\a_\pm \, e^{-\frac{i\pi}4}}{\sqrt2} \frac{\hh\,\hq\,(1-e^{-i\pp})}{\sqrt{e_\pm+1 \pm 2 \hh\,\q\,\sin\frac\pp2}} \ , &
\ \ \ d_\pm =\, & \frac{\a_\pm^{-1}e^{\frac{i\pi}4}}{\sqrt 2} \sqrt{e_\pm+1\pm 2\hh \, \q\,\sin \tfrac\pp2}
\ , \label{ppm}
\end{align}
where, as before, $\a_\pm$ are free phases, that are allowed to depend on the momentum. In the BMN limit \rf{bmn}, setting $\a_\pm = 1 + \mathcal{O}(\hh^{-1})$,
these expressions reduce to the corresponding ones in \eqref{Functions}.

Substituting $M_\pm$, $C_\pm$, $P_\pm$ and $K_\pm$ from \eqref{uq} into the shortening conditions \eqref{bhshort}
leads to the following exact dispersion relation \cite{htb}
\begin{equation}\begin{split}\la{iq}
e_\pm^2 =\, & (1 \pm 2 \hh \, \sig\, \sin \frac \pp2)^2 + 4 \, \hh^2 \,\hq^2 \, \sin^2\frac \pp2
\\ =\, & 1 \pm 4 \hh \, \sig\, \sin \frac \pp2 + 4 \, \hh^2 \, \sin^2 \frac \pp2 \ .
\end{split}\end{equation}

\renewcommand{\theequation}{4.\arabic{equation}}
\setcounter{equation}{0}
\section{Exact dispersion relation}\label{sec4}

Let us now discuss some features of the exact dispersion relation \rf{iq} or, for a positive-energy particle state,
\be
e_\pm = \sqrt{\hq^2 + (2 \, \hh \, \sin \frac \pp2 \pm \sig)^2 } \ . \la{ds} \ee

\subsection{General structure and limits}

Assuming the relation \rf{ka} between $\hh$ and the quantized coefficient $k$ of the bosonic WZ term in the string action
(related to the NS5-brane charge), this dispersion relation can be written also as
\be \la{dss}
e_\pm = \sqrt{ 1 \pm {2k \ov \pi} \sin \frac \pp2 +\ 4 \, \hh^2 \, \sin^2 \frac \pp2 } \ .
\ee
Eq.\rf{ds} interpolates between the RR case ($\hq=1, \ q=0$) when it becomes the standard
magnon dispersion relation and the NSNS case ($\hq=0, \ q=1$).
In the latter case when the world-sheet theory becomes
the superstring generalization of the $SU(1,1) \times SU(2)$ WZW theory with level $k$,
eq.\rf{ds} reduces to (cf. \rf{ka})\foot{A similar ``massless'' 
dispersion relation $ e= 2 h \big| \sin {p\ov 2}\big| $ appeared already in \ci{b1}
in the limit $\alpha \to 0$ or $\alpha \to 1$
of the $AdS_3 \times S^3 \times S^3 \times S^1$ theory (with
the radii
of the two spheres parametrized as $R_1^2 = \alpha^{-1}, \ R_2^2 = ( 1 - \alpha)^{-1}$).\fnsv}
\be e_\pm =\big| 1 \pm 2\hh\,\sin\frac \pp2 \big| \ , \ \ \ \ \ \qquad \ \ \ \ \ \ \hh={ k \ov 2 \pi} \ . \la{wzw} \ee
For small $\pp$ one therefore finds a massless dispersion relation,
in agreement with the world-sheet expectations
(in the BMN limit \rf{get} we get $e_\pm = |p \pm 1|$).
As discussed below, \rf{wzw} can be interpreted as corresponding to a lattice analog of a massless
chiral scalar or fermion kinetic operator.

In the BMN limit \rf{bmn}
the relation \rf{ds} reduces to
\begin{equation}
e_\pm = \sqrt{ \hq^2 + (p \pm \sig)^2 } + \mathcal{O}(\hh^{-2}) \ , \la{get}
\end{equation}
matching the result \rf{1_5} following directly from the string world-sheet perturbation theory \cite{bmn,cz,htb}.
In the different strong coupling limit -- semiclassical or ``giant magnon'' \ci{hma} limit
when $\pp$ stays finite while $\hh \to \infty $, eq.\rf{ds} reduces to
\begin{equation}\la{gia}
e_\pm = 2 \hh \, \sin \frac{\pp}{2} \pm \sig\ + \mathcal{O}(\hh^{-1}) \ .
\end{equation}
This implies that the classical energy (minus the angular momentum) of the corresponding
``giant magnon'' solution
should not dependent on $\q$,
i.e. on the value of the NSNS flux. At the same time,
there should be a string 1-loop correction proportional to $\q $
(indeed, there was no 1-loop correction in the $\q = 0$
case \cite{ps,david}).
It would be interesting to confirm the presence of this correction
by a string-theory computation of the
one-loop correction to the corresponding
giant-magnon energy, thus providing a non-trivial
check of the exact dispersion relation \rf{ds}.

As there is little solid knowledge about the corresponding dual 2-d CFT (beyond the supersymmetry protected BPS states and moduli space, 
cf. \ci{ras} and references there) it is hard to comment on the possible meaning of \rf{ds} or \rf{dss} in the small string tension or
weak coupling region $\hh \to 0 $.
In general, the identification of the parameter $\hh$ in \rf{ds} with the string tension $\sql \ov 2\pi$
in \rf{ef} may be true only in the strong-coupling limit $\sql \gg 1$, i.e. $\hh$ may be a non-trivial function of $\l$.
This finite
renormalization appears to be absent in the pure RR case of $\q = 0$,
and it should also be absent in the pure NSNS case of $\q = 1$
when $\hh$ is directly related to the integer
level $k$ (cf. \rf{ka}). However, it may be present for a generic value of $\q$.
Indeed, there is a 1-loop shift in $\hh(\l)$ in the case of another 1-parameter deformation
of the $AdS_3 \times S^3 \times T^4$ theory -- the $AdS_3\times S^3 \times S^3 \times S^1$ theory \ci{ab,bec}.
It would be important to investigate this by a direct 1-loop superstring computation for $\q \neq 0 $.

Let us note that from the world-sheet sigma model point of view, the scattering of states with non-trivial
dispersion relation \rf{iq} or \rf{ds} corresponds to the scattering of solitonic ``giant-magnons''
which may be viewed as elementary massive light-cone gauge quanta (usual BMN ``magnons'')
``dressed'' by quantum corrections to all orders in the $\hh^{-1}$ expansion. The fact that the exact
S-matrix of the elementary excitations with the standard quadratic relativistic dispersion relation \rf{get}
can be rewritten as an S-matrix for the scattering of such ``dressed'' states with the dispersion relation \rf{iq} is, of course,
a non-trivial consequence of the integrability of the model.
This should apply also to the special $q=1$ case when
the world-sheet theory is described by the WZW model:
here the unfamiliarly looking dispersion relation \rf{wzw} should correspond again to an analog of the
``giant magnon''
soliton in the classical WZW theory.

\def \ppp {\pp}
\def \ha {{1 \ov 2}}
\def \ry {{\rm y}}
\def \rn {{\rm k}}
\def \L {{\ell}}
\def \ve {{\varepsilon}}
\def \J {{\cal J}}
\def \rn {{\td n}}
\subsection{Lattice origin of the dispersion relation}

Let us now show that
the exact dispersion relation \rf{ds} or \rf{dss} corresponds to a discretization
(in the spatial world-sheet direction) of the second-order differential operator appearing in the quadratic part of the $q\not=0$
$AdS_3 \times S^3$ string action
expanded near the BMN geodesic
(or, equivalently, written in the BMN light-cone gauge).
It is thus a natural 1-d lattice (or ``spin chain'') analog of the BMN dispersion relation \rf{get}.

As was discussed in \ci{htb},
the quadratic term in the bosonic string action expanded near the BMN geodesic has the following form
\be
I & = & \ha \hh \int d \tau d\sigma \Big(
- \del^a y_r \del_a y_r - y_r y_r +\ q\, \epsilon_{rs} y_r \del_1 y_s \Big) \no
\\ & = & \ha \hh \int d \tau d\sigma \Big( \dot y_r^2 - y'^2_r - y_r^2 +\ 2q\, y_1 y'_2 \Big) \ . 
\la{ys} \ee
Here the $q$-dependent term originates from the WZ term or the $B$-field coupling.
$y_r $ ($r,s=1,2$) are two real scalars representing the transverse fluctuations in $S^3$.
The same action is found for
the two scalars $z_r$ representing the transverse fluctuations in $AdS_3$.
The massive fermionic modes
have (after ``squaring'') an equivalent kinetic operator, albeit with the
mass term $\sim \hq$ originating not from the curvature as for the bosons, but rather from the RR flux coupling.
The
corresponding dispersion relation is then given by the leading term in \rf{get}.

Let us now assume that the spatial direction $ \s$ is compact with length
$\L= 2\pi \J$
(we rescale $\tau$ and $\s$ by the semiclassical $S^3$ angular momentum or rotation frequency parameter $\J$).\foot{Note that for the decompactified $\s$ case, corresponding to the limit of $\J \to \infty$,
the $q$-dependence
of the derivative term in \rf{ys}
can be eliminated by a local rotation $y_1 + i y_2 = e^{i q \s} v(\tau, \sigma)$
or, equivalently, by a shift of the continuous momentum $p$ in \rf{get}. This is no longer possible for finite $\J$,
unless $q \J$ is an integer to ensure that $v$ is still periodic in $\s$.\fnsv}
Furthermore, let us also discretize $\s$ into $J$ points with step $\ve$,
\be \la{ay}
\ve = {\L\ov J} = 2 \pi {\J\ov J} \ , \ \ \ \ \ \ \ \ \ y_{r (n)} (\tau) = y_r (\tau, n\ve ), \ \ \ \ \ \ \ n=0, ..., J-1, \ \ \ \ \ \ y_{r(J)}= y_{r(0)} \ . \ee
Assuming that the spatial derivative is defined as
$y'_r \to \ve^{-1} ( y_{r(n+1)} - y_{r(n)} ) , $
the discrete version of the action \rf{ys} becomes
\be
I = \ha \hh \ve \sum_{n=1}^{J} \int d \tau \Big[
\dot y_{r(n)}^2 - \ve^{-2} ( y_{r (n+1)} - y_{r (n)})^2 - y_{r(n)}^2
+ 2q\, y_{1 (n)} ( y_{2 (n+1)} - y_{2 (n)}) \Big] \ .
\la{yss} \ee
The corresponding equations of motion for $y_{1(n)}$ and $y_{2(n)}$ are then
\be
&&
\ddot y_{1(n)} + y_{1(n)} - \ve^{-2} ( y_{1 (n+1)} - 2 y_{1 (n)} + y_{1 (n-1)}) - 2 q \ve^{-1} ( y_{2 (n+1)} - y_{2 (n)}) =0 \ , \\
&&
\ddot y_{2(n)} + y_{2(n)} - \ve^{-2} ( y_{2 (n+1)} - 2 y_{2 (n)} + y_{2 (n-1)}) + 2 q \ve^{-1} ( y_{1 (n )} - y_{1(n-1)}) =0 \ . \la{eqs}
\ee
As in the standard Klein-Gordon operator case of $q=0$ (see, e.g., \ci{st,paw}) these can be solved
using the momentum space eigenfunctions, i.e. by replacing
\be
y_{r(n)}(\tau) \ \to\ \ry_r e^{ - i e\tau } e^{ - i \ppp n } \ , \ \ \ \ \ \ \ \ \ \ \
\ppp = { 2\pi \rn \ov J} \ , \ \ \ \ \ \ \rn
=0, 1, ..., J-1 \ . \la{ps}\ee
This gives
\be
&& ( - e^2 +1 + 4 \ve^{-2} \sin^2 {\pp \ov 2} ) \ry_1 - 2 q \ve^{-1} (e^{-i \pp} - 1 ) \ry_2 =0 \ ,\no \\
&& ( - e^2 +1 + 4 \ve^{-2} \sin^2 {\pp \ov 2} ) \ry_2 + 2 q \ve^{-1} ( 1 - e^{i \pp} ) \ry_1 =0 \ .\la{eqq}
\ee
The corresponding dispersion relation is
$
( e^2 -1 - 4 \ve^{-2} \sin^2 {\pp \ov 2} )^2 - 16 q^2 \ve^{-2} \sin^2 {\pp \ov 2} =0,$ or
\be e^2_\pm = 1 + 4 \ve^{-2} \sin^2 {\pp \ov 2} \pm 4 q \ve^{-1} \sin {\pp \ov 2} \ , \la{spe}
\ee
which is equivalent to the exact dispersion relation in \rf{iq} upon making the following identification
\be \la{rp}
\ve^{-1} = \hh \ .
\ee
Thus the step of the lattice $\ve$ has the interpretation of the inverse of string tension.
Then using \rf{ef},\rf{ay} we conclude that $ J= \sql \J$, which is the familiar relation between the
semiclassical $\J$ and exact $J$ angular momentum.

In the special case of $q=1$ the dispersion relation \rf{ds} simplifies to \rf{wzw}.
In this case the RR flux is absent and the string theory is described by
supersymmetric generalization of the $SU(1,1) \times SU(2)$ WZW model and should therefore have a massless
perturbative spectrum. This is indeed so in the ``non-compact'' ($\J\to \infty$)
or small $\pp$ limit of \rf{wzw} or \rf{spe}.
The exact relation \rf{wzw} also has a discrete
interpretation ---
as the dispersion relation for the ``left'' and ``right''
massless operators $\del_0 \pm \del_1$ on a 1-d spatial lattice.
Indeed, the chirality of the WZW equations of motion imply that, to quadratic fluctuation order,
they reduce to the equations for the ``left'' and ``right'' chiral scalars. Equivalently,
one may start with a massless fermionic Lagrangian\,\foot{Here $i$ stands for a chemical potential term
leading to a gap in the energy to match \rf{spe}.\fnsv}
\be
\la{fi} L= \psi^* ( \del_0 \pm \del_1 - i ) \psi \ . \ee
Starting with \rf{fi} and discretizing $\s$ as discussed above, we find the dispersion relation
\be
e_\pm = \big|1 \pm 2 \ve ^{-1}\sin {\ppp\ov 2} \big|
\ , \la{eew} \ee
which is indeed the same as \rf{wzw} provided we assume \rf{rp}.

We conclude that the NSNS 3-form flux (or $q$-dependent) term in the generalization \rf{iq} of the BMN dispersion relation
has a natural discrete (1-d lattice) origin. Heuristically, this suggests that for fixed $J$ the world sheet
becomes effectively discrete at finite string tension. Note that in the $AdS_5 \times S^5$ context
the relation between the spin chain description of the SYM dilatation operator at weak coupling
(in the ``quadratic'' approximation, i.e. ignoring scattering of magnons)
and the quadratic BMN term in the string action can be made precise by including all higher-derivative terms
in the Landau-Lifshitz-type description of fluctuations near the ferromagnetic vacuum (see \ci{ryt}).

Let us note that as was recently discussed in \ci{paw}, the Casimir vacuum energy of a 2-d scalar field of mass 1 on
a periodic spatial lattice of $J$ points with step $\ve= \hh^{-1}$ (with the standard
dispersion relation given by the $q=0$ limit of \rf{spe})
can be identified with the free energy of a gas of particles on an infinite line
with temperature $1/J$ and the ``mirror'' energy $\td e$ \ci{am,afa}
\be
E_{\rm vac} (J,\hh)= \int^\infty_0 { d \td \pp \ov \pi} \log \big[1 - e^{-J \td e (\td \pp, \hh) }\big] \ , \ \ \ \ \qquad
\td e = 2\, {\rm arcsinh}\, {\sqrt {1 + \td \pp^2} \ov 2 \hh} \ . \la{mi}
\ee
Here $\td e, \td \pp$ are related to $e,\pp$ in the original dispersion relation
$e^2 = 1 + 4 \hh^2 \sin^2 { \pp \ov 2}$
by the double-Wick rotation, $e\to i \td \pp, \ \pp \to i \td e$.
A similar expression should appear also in the $q\not=0$ case, where eq.\rf{iq} or \rf{spe},\rf{rp} implies that the
corresponding mirror dispersion relation is ($\hq^2 = 1 - q^2 $)
\be
\td e_\pm = 2\, {\rm arcsinh} \, {\sqrt {\hq^2 + \td \pp^2} \mp i q \ov 2 \hh} \ . \la{mir}
\ee
The analog of \rf{mi} will contain the sum of two $\log$ terms corresponding to the two signs in \rf{mir}.
As was shown in \ci{paw}, for large $\J$ and large or small $\hh$
the energy $E_{\rm vac} $ in \rf{mi}
has either $e^{-\J}$ (``Luscher'') or $\hh^{2J}$ (``wrapping'') behaviour.
It would be of interested to study how this is modified in the $q\not=0$ case.
In particular, the $q=1$ case appears to be very special.

\renewcommand{\theequation}{5.\arabic{equation}}
\setcounter{equation}{0}
\section{Exact S-matrix}\label{sec5}

Let us now turn to the question of the exact generalization of the tree-level expression \rf{fu} for the S-matrix
\rf{bct}.

\subsection{$q=0$ case}\label{qzero}

Let us start by reviewing the pure RR case discussed in \ci{bssst} (see also \ci{bsz,ahn,bor1} for related
earlier work).
As was briefly discussed in section \ref{sec21}, for $q=0$
there is a formal symmetry under the
interchange of the two irreducible atypical representations.
Therefore, for the remainder of this subsection we may drop the subscripts $\spm$ on
both the representation parameters and the parametrizing functions of the S-matrix.

For $\q = 0$
the standard relations between the Zhukovsky variables $x^\pm = x^\pm (\pp)$
and the energy and momentum are
as in the \adss case \ci{beis1}\,\foot{Note that the coupling $\hh$ used here is twice the coupling $h$ used
in \cite{bssst}.\fnsv}
\bea \la{xx}
&& e^{i\pp} = \frac{x^+}{x^-} \ , \qquad\qquad
e +1 = i \hh (x^- - x^+) \ ,
\eea
In these variables the dispersion relation \eqref{apiii} takes the following familiar form
\be x^+ + \frac{1}{x^+} -x^- - \frac{1}{x^-} = \frac{2i}{\hh} \ , \la{yy} \ee
and solving for $x^\pm$ in terms of $e$ and $\pp$ we find
\begin{equation}\begin{split}
x^\pm = r\, e^{\pm { \frac{i\pp}{2}}} \ , \ \ \ \ \ \ \ \ \ \ \ \quad
r= \frac{e + 1 }{2 \hh\,\sin\frac \pp2} =\frac{2\hh\,\sin\frac\pp2}{e - 1 } \ .
\label{4444}
\end{split}\end{equation}
The representation parameters $a,b,c$ and $d$ in \rf{ad},\rf{bep} can then be written as\,\foot{Using
the mapping in footnote \ref{footcompare1}, we see that for exact agreement with \cite{bssst}
we should set $\a = e^{\frac{i\pi}{4}}\sqrt{\frac{x^+}{x^-}}$.\label{footcompare2}\fnsv}
\be
&& a =\, \a \,e^{-\frac{i\pi}4}\sqrt{\tfrac{\hh}2}\ \eata \ , \qquad \qquad
b =\, \a^{-1} e^{-\frac{i\pi}4} \sqrt{\tfrac{\hh}2} \ \frac{\eata}{x^-} \ , \no \\
&& c =\, \a\, e^{\frac{i\pi}4} \sqrt{\tfrac{\hh}2} \ \frac{\eata}{x^+} \ , \qquad \qquad
d =\, \a^{-1}e^{\frac{i\pi}4} \sqrt{\tfrac{\hh}2}\ \eata \ , \label{da}
\ee
where
\begin{equation}\la{ett}
\eata \equiv \sqrt{i(x^- - x^+)} = \sqrt{\frac{e+1}{\hh}}\ .
\end{equation}

The functions parametrizing the exact S-matrix \eqref{bct} are given by \cite{bssst}\,\footnote{Here we have
written the parametrizing functions of the S-matrix in the so-called ``string frame'' (see, e.g., \ci{an}). Therefore,
it is this frame that one should use to compare to \cite{bssst}. The transformation from
the ``spin-chain frame'' to the ``string frame'' is given explicitly in appendix E of \cite{bor1}.
The mapping between the functions parametrizing the S-matrix
used here and those used in \cite{bssst,bor1} is then given by
(see also footnote \ref{foot2})
\begin{equation*}\begin{split}
& \So A\big|_{\ga=0} = S \sqrt{\frac{x^+\xpr^-}{x^-\xpr^+}} \ ,\qquad \qquad \Sio \bar A\big|_{\ga=0} = \tilde S\sqrt{\frac{\xpr^-}{\xpr^+}}\ \sqrt{\frac{1 - \frac{1}{x^-\xpr^-}}{1 - \frac{1}{x^+\xpr^+}}} \ ,
\\ & L_1 = A^{LL} = A^{RR} \ , \quad \Lambda_1 = - F^{LL} = - F^{RR} \ ,\quad L_3 = B^{LL} = B^{RR} \ , \quad
\Lambda_3 = D^{LL} = D^{RR} \ ,
\\ & L_5 = C^{LL} = C^{RR} \ , \quad \Lambda_5 = E^{LL} = E^{RR} \ , \quad L_6 = C^{LR} = C^{RL} \ , \quad \Lambda_6 = D^{LR} = D^{RL}\ ,
\\ & L_2 = A^{LR} = A^{RL} \ , \quad \Lambda_2 = - E^{LR} = -E^{RL} \ , \quad L_4 = -B^{LR} = -B^{RL} \ , \quad \Lambda_4 = F^{LR} = F^{RL} \ .
\end{split}\end{equation*}
Furthermore, to establish exact agreement the phase $\a$
in \eqref{da}
(which just parametrizes the normalization of the fermionic states with respect to the bosonic ones)
should be set to
$e^{\frac{i\pi}{4}}\sqrt{\frac{x^+}{x^-}}$, in concord with footnote \ref{footcompare2}.\label{footcompare3}\fnsv}
\begin{align}\no
L_{1} = \ & \So{} \ , &
\Lambda_{1} = \ & \So{} \sqrt{\frac{x^-\xpr^+}{x^+\xpr^-}}\,
\frac{x^+ - \xpr^-}{x^- - \xpr^+} \ ,
\\\no
L_{3} = \ & \So{} \sqrt{\frac{x^-}{x^+}} \, \frac{x^+ - \xpr^+}{x^- - \xpr^+} \ , &
\Lambda_{3} = \ & \So{} \sqrt{\frac{\xpr^+}{\xpr^-}}\, \frac{x^- - \xpr^-}{x^- - \xpr^+} \ ,
\\\no
L_{5} =\ & -i\,\frac{\a}{\a'} \,\So{} \sqrt{\frac{x^-\xpr^+}{x^+\xpr^-}}\,
\frac{\eata\eata'}{x^- - \xpr^+} \ , &
\Lambda_{5} =\ & -i\,\frac{\a'}{\a} \, \So{} \frac{\eata\eata'}{x^- - \xpr^+} \ ,
\\\no
L_{6} = \ & \Sio{} \ , &
\Lambda_{6} = \ & \Sio{} \sqrt{\frac{x^-\xpr^-}{x^+\xpr^+}}\, \frac{1-x^+\xpr^+}{1-x^-\xpr^-} \ ,
\\\no
L_{2} = \ & \Sio{} \sqrt{\frac{x^-}{x^+}}\, \frac{1-x^+\xpr^-}{1-x^-\xpr^-} \ , &
\Lambda_{2} = \ & \Sio{} \sqrt{\frac{\xpr^-}{\xpr^+}}\, \frac{1-x^-\xpr^+}{1-x^-\xpr^-} \ ,
\\\la{ses}
L_{4} = \ & i\, \a\a'\, \Sio{} \sqrt{\frac{x^-\xpr^-}{x^+\xpr^+}}\, \frac{\eata \eata'}{1-x^-\xpr^-} \ ,&
\Lambda_{4} = \ & i\,\frac{1}{\a\a'}\, \Sio{} \frac{\eata \eata'}{1-x^-\xpr^-} \ .
\end{align}
Here the primed kinematic variables correspond to primed fields in \rf{bct}
and
the phases $A,\bar A$ \eqref{pha}, which contain the dependence on the gauge parameter $\ga$, take the usual exact form \cite{afr}
\begin{equation}
A = \bar A = \exp\big[-{\textstyle \frac i2}(\ga-\tfrac12)(e' \pp - e \pp')\big] \ , \la{57}
\end{equation}
which can also be written explicitly in terms of $x^\pm$ using \rf{xx}.

Crucially, this S-matrix is completely fixed, up to the two phases $\So$ and $\Sio$, just by demanding
the invariance under the four supercharges. In particular, one does not need to impose the dispersion relation \eqref{yy}.
Furthermore, it can be checked that this S-matrix satisfies the Yang-Baxter equation, QFT unitarity, and also braiding unitarity
so long as the phases satisfy \cite{bssst}
\begin{equation}\begin{split}
\So(x^+,x^-;\xpr^+,\xpr^-)\,\So(\xpr^+,\xpr^-;x^+,x^-) = & 1 \ ,
\\\Sio(x^+,x^-;\xpr^+,\xpr^-)\,\Sio(\xpr^+,\xpr^-;x^+,x^-)
= & \sqrt{\frac{x^+ \xpr^+}{x^-\xpr^-}}\, \frac{1-x^-\xpr^-}{1-x^+\xpr^+}\ .\la{pha2}
\end{split}\end{equation}
Again this does not require the use of the dispersion relation.

\

This suggests that a natural strategy to generalize to the case of $\q \neq 0$
would be to find a parametrization such that the representation parameters
$a_\pm,b_\pm,c_\pm$ and $d_\pm$ \eqref{ppm} take the same form as in the $q = 0$ case \eqref{da}
when written in terms of the corresponding Zhukovsky variables
$x^\pm_\spm$ (the subscripts correspond to the positive/negative charged states).
While this will modify the dispersion relation \eqref{yy},
the S-matrix will remain unchanged up to the introduction of the $\spm$ subscripts on $x^\pm$ and $\xpr^{\pm}$.
Due to the block-diagonal nature of the S-matrix (the two-particle states $|++\ket$
always scatter into ${ |++\ket}$ states and similarly for $|+-\ket$, $|-+\ket$ and $|--\ket$)
the introduction of the subscripts
on the Zhukovsky variables should not affect the satisfaction of the Yang-Baxter equation
or braiding unitarity. Furthermore, if
the conjugation of the Zhukovsky variables, $(x^\pm)^* = x^\mp$ is simply generalized to
$(x^\pm_\spm)^* = x^\mp_\spm$, then the QFT unitarity property of the S-matrix 
should still hold.

\

Before turning to a detailed discussion of the $q\not=0$ case let us
make a brief comment on crossing symmetry. Choosing $\a$ to have the form\,\footnote{To recall, $\a$ is a phase parametrizing the normalization of the fermionic states with respect to the bosonic ones, which can depend on the momentum. There are various conventions used in the literature, which, in general, can be written in the form given in eq.\eqref{form}.\fnsv}
\begin{equation}\label{form}
\a = e^{{i\g}\ov 2}\, \big(\frac{x^+}{x^-}\big)^\beta\ ,
\end{equation}
where $\b$ and $\g$ are arbitrary real numbers,
the S-matrix \eqref{bct}, \eqref{ses} has a crossing symmetry \cite{bssst}, so long as the two phases in \rf{ses} are related in the following way:
\begin{equation}\begin{split}
\So^c = \Sio \sqrt{\frac{x^-}{x^+}}\, \frac{1-x^+\xpr^-}{1-x^-\xpr^-}\ , \qquad\qquad
\Sio^c = \So \sqrt{\frac{x'^+}{x'^-}}\, \frac{x^- - \xpr^-}{x^--\xpr^+}\ .\label{cross1}
\end{split}\end{equation}
Here the label ${}^c$ denotes that the corresponding
arguments are taken as $(\bar x'{}^+,\bar x'{}^-; x^+,x^-)$ instead of original $( x^+,x^-; x'{}^+, x'{}^-)$
where the ``crossed'' Zhukovsky variables
$\bar x^\pm$ are, as usual, given by
\begin{equation}
\label{czv}
\bar x^\pm = \frac 1{x^\pm} \ ,
\end{equation}
corresponding to $\bar e = - e \ , \ \bar \pp = - \pp$.
The crossing symmetry
can be seen at the level of the parametrizing functions \eqref{ses}
from the following identities
\begin{align}
L_1^c = & L_2\ ,
&
\Lambda_1^c= & \Lambda_2 \ ,
&
L_3^c = & \Lambda_6 \ ,
&
\Lambda_3^c = & L_6\ ,
&
L_5^c = & - i e^{i\g} \Lambda_4 \ ,
&
\Lambda_5^c = & - i e^{-i\g} L_4 \ ,\nonumber
\\
L_2^c = & L_1 \ ,
&
\Lambda_2^c = & \Lambda_1 \ ,
&
\Lambda_6^c = & L_3 \ ,
&
L_6^c = & \Lambda_3\ ,
&
\Lambda_4^c = & i e^{-i\g} \Lambda_5\ ,
&
L_4^c = & i e^{i\g}L_5\ .
\label{cross2}
\end{align}

\subsection{\texorpdfstring{$q\neq0$}{q!=0} case}\label{qneq1}

To generalize to the $\q \not =0$ case let us
first modify the relations between the Zhukovsky variables $x^\pm$ and $e,\pp$ \eqref{xx} as
follows:\footnote{To be completely
general, one could work in terms of the representation parameters $C_\pm,M_\pm$ and $U_\pm$, where
$P_\pm = \ha {\tilde \hh}(1-U_\pm^2)$ and
$K_\pm = \ha {\tilde \hh}(1-U_\pm^{-2})$ \ (${\tilde \hh}$ is some proportionality coefficient, cf.\rf{uq}, that should go like $\hh\, \hq$ for large $\hh$)
are fixed by requiring that the coproduct should equal its opposite for the central extensions.
Defining
\begin{equation*}
U_\pm^2 = \frac{x_\spm^+}{x_\spm^-} \ , \qquad\qquad
2C_\pm + M_\pm = i \tilde \hh \, (x_\spm^- - x_\spm^+) \ ,
\end{equation*}
the dispersion relation then takes the following form
\begin{equation*}
x_\spm^+ + \frac1{x_\spm^+} - x_\spm^- - \frac1{x_\spm^-} = \frac{2iM_\pm}{\tilde\hh}\ .
\end{equation*}
The rest of this subsection then proceeds almost identically (only the definitions of $x_\spm^\pm$ in terms of $e_\pm$ and $\pp$ \eqref{444}
and the discussion of the semiclassical limit at the end are modified)
so long as the representation parameters have the correct near-BMN expansion \eqref{Functions}.
Furthermore, the discussion in subsection \ref{qeq1} also applies assuming that
$\tilde \hh \sim \hq$ as $q \to 1$.\label{footdis}\fnsv}
\bea \la{xxq}
&& e^{i\pp} = \frac{x_\spm^+}{x_\spm^-} \ , \qquad\qquad
e_\pm + 1 \pm 2\, \hh \,q\, \sin \frac \pp2 = i \hh\,\hq\, (x_\spm^- - x_\spm^+) \ .
\eea
Written in terms of these new variables $x_\spm^\pm$ the dispersion relation
\eqref{iq} takes the following form\foot{A different 
modification of the dispersion relation for $x^\pm$ appeared in the $AdS_3 \times S^3 \times S^3\times S^1$ case \ci{b1}. In that case
the modes were split into two halves, one with mass $\alpha$ and the other with mass $1-\alpha$.
The modification then amounted to effectively rescaling the $\hh^{-1}$ term in the r.h.s. of the analog of \rf{4_7}, or, equivalently,
taking $M_\pm = \a$ or $M_\pm=1-\a$ in footnote \ref{footdis}. This is in contrast to the situation here, for which the $q$-dependent modification
depends on the momentum and requires $M_+ \neq M_-$.
Note also that the $q\to 1$ limit of the expression \rf{4_7} should be taken with care as $x_\spm^\pm \sim \hq^{-1}$ or $\hq$.
This is discussed further in section \ref{qeq1}.\fnsv}
\be\label{4_7}
\hq \,\Big(x_\spm^+ + \frac1{x_\spm^+} - x_\spm^- - \frac1{x_\spm^-}\Big) \pm 2 q \Big(\sqrt{\frac{x_\spm^-}{x_\spm^+}} - \sqrt{\frac{x_\spm^+}{x_\spm^-}}\Big) = \frac{2i}{\hh}\ .
\ee
Solving for $x^\pm_\spm$ in terms of $e_\pm$ and $\pp$ for general $q$ we find the following generalization
of \eqref{4444}
\begin{equation}\begin{split}
x_\spm^\pm = r_\spm\, e^{ \pm \frac{i\pp}{2}}\ , \qquad \ \ \
r_\spm = \frac{e_\pm + 1 \pm 2\hh\,q \,\sin \frac \pp2}{2 \hh\,\hq\,\sin\frac \pp2} =
\frac{2\hh\,\hq\,\sin\frac\pp2}{e_\pm - 1 \mp 2\hh\,q \,\sin \frac \pp2}
\ .\label{444}
\end{split}\end{equation}
While the dispersion relation \eqref{4_7} is a non-trivial modification of the $q=0$ relation
\eqref{yy}, the expressions for the representation parameters $a_\pm,b_\pm,c_\pm$ and $d_\pm$ \eqref{ppm} in terms
of $x^\pm_\spm$ are found to have the same form as in 
\eqref{da}, up to a universal rescaling by $\sqrt{\hq}$, i.e.
\begin{align}
a_\pm =\, & \a_\pm \,e^{-\frac{i\pi}4}\sqrt{\tfrac{\hh\,\hq}2}\ \eata_{_\pm} \ , &\
b_\pm =\, & \a_\pm^{-1} e^{-\frac{i\pi}4} \sqrt{\tfrac{\hh\,\hq}2} \ \frac{\eata_{_\pm}}{x_{_\pm}^-} \ ,\no \\
c_\pm =\, & \a_\pm\, e^{\frac{i\pi}4} \sqrt{\tfrac{\hh\,\hq}2} \ \frac{\eata_{_\pm}}{x_{_\pm}^+} \ , &
d_\pm =\, & \a_\pm^{-1}e^{\frac{i\pi}4} \sqrt{\tfrac{\hh\,\hq}2}\ \eata_{_\pm} \ . \label{daq}
\end{align}
Here, like in \rf{ett},
\begin{equation}\la{etq}
\eata_{_\pm} \equiv \sqrt{i(x_{_\pm}^- - x_{_\pm}^+)} = \sqrt{\frac{e_\pm+1\pm 2\hh\,q\,\sin \frac \pp2}{\hh\,\hq}}\ .
\end{equation}
As this rescaling by $\hq$ does not affect the action of the
supersymmetry algebra,
we can immediately write down the functions
parametrizing the exact S-matrix \eqref{bct} in the case of $q\not=0$
by generalizing the expressions in \rf{ses} as follows
\begin{align}\no
L_{1_\pm} = \ & \So{}_{_\pm} \ , &
\Lambda_{1_{\pm}} = \ & \So{}_{_\pm} \sqrt{\frac{x_\spm^-\xpr_\spm^+}{x_\spm^+\xpr_\spm^-}}\,
\frac{x_\spm^+ - \xpr_\spm^-}{x_\spm^- - \xpr_\spm^+} \ ,
\\\no
L_{3_{\pm}} = \ & \So{}_{_\pm} \sqrt{\frac{x_\spm^-}{x_\spm^+}} \, \frac{x_\spm^+ - \xpr_\spm^+}{x_\spm^- - \xpr_\spm^+} \ , &
\Lambda_{3_{\pm}} = \ & \So{}_{_\pm} \sqrt{\frac{\xpr_\spm^+}{\xpr_\spm^-}}\, \frac{x_\spm^- - \xpr_\spm^-}{x_\spm^- - \xpr_\spm^+} \ ,
\\\no
L_{5_{\pm}} =\ & -i\,\frac{\a_\pm}{\a'_\pm} \,\So{}_{_\pm} \sqrt{\frac{x_\spm^-\xpr_\spm^+}{x_\spm^+\xpr_\spm^-}}\,
\frac{\eata_\spm\eata_\spm'}{x_\spm^- - \xpr_\spm^+} \ , &
\Lambda_{5_{\pm}} =\ & -i\,\frac{\a'_\pm}{\a_\pm} \, \So{}_{_\pm} \frac{\eata_\spm\eata_\spm'}{x_\spm^- - \xpr_\spm^+} \ ,
\\\no
L_{6_{\pm}} = \ & \Sio{}_{_\pm} \ , &
\Lambda_{6_{\pm}} = \ & \Sio{}_{_\pm} \sqrt{\frac{x_\spm^-\xpr_\smp^-}{x_\spm^+\xpr_\smp^+}}\, \frac{1-x_\spm^+\xpr_\smp^+}{1-x_\spm^-\xpr_\smp^-} \ ,
\\\no
L_{2_\pm} = \ & \Sio{}_{_\pm} \sqrt{\frac{x_\spm^-}{x_\spm^+}}\, \frac{1-x_\spm^+\xpr_\smp^-}{1-x_\spm^-\xpr_\smp^-} \ , &
\Lambda_{2_{\pm}} = \ & \Sio{}_{_\pm} \sqrt{\frac{\xpr_\smp^-}{\xpr_\smp^+}}\, \frac{1-x_\spm^-\xpr_\smp^+}{1-x_\spm^-\xpr_\smp^-} \ ,
\\\la{sqs}
L_{4_{\pm}} = \ & i\, \a_\pm\a'_\mp\, \Sio{}_{_\pm} \sqrt{\frac{x_\spm^-\xpr_\smp^-}{x_\spm^+\xpr_\smp^+}}\, \frac{\eata_\spm \eata_\smp'}{1-x_\spm^-\xpr_\smp^-} \ ,&
\Lambda_{4_{\pm}} = \ & i\,\frac{1}{\a_\pm\a'_\mp}\, \Sio{}_{_\pm} \frac{\eata_\spm \eata_\smp'}{1-x_\spm^-\xpr_\smp^-} \ .
\end{align}
As was already mentioned in section \ref{qzero}, since the S-matrix at $q=0$ satisfies the Yang-Baxter equation without the need
to use
the dispersion relation, and since it has a block-diagonal structure, the above generalization of this
S-matrix to $q \neq 0$ case should
still satisfy the YBE.
Indeed, we have checked explicitly that this is the case for the S-matrix \rf{bct} with \rf{sqs}.

Let us recall that the tree-level S-matrix's generalization to non-zero $q$ \cite{htb}, summarized in section
\ref{sec1}, was remarkably simple. In particular, the functions $l_{1,2,3}$ only depend on $q$ through the
dispersion relation. This simplicity is apparent in the exact S-matrix when written in terms of the Zhukovsky variables.
It is worth noting that if the exact S-matrix is written in
terms of the energy and momentum variables 
 this simplicity is no longer manifest due to the non-trivial all-order 
definition of    the Zhukovsky variables \eqref{444}
(in particular, their  dependence on $q$ is not only through the dispersion relation).
This is hinted at in the tree-level results by the more complicated structure of $l_{4,5}$ compared to $l_{1,2,3}$ \eqref{lel}. 

The phases $A,\bar A$ in \eqref{57}, which contain the dependence on the gauge parameter $\ga$, have the
following natural generalization to $q\neq 0$
\begin{equation}\la{19}
A_{_\spm} = \exp\big[-{\textstyle \frac i2}(\ga-\tfrac12)(e_\pm' \pp - e_\pm \pp')\big] \ , \qquad
\bar A_{_\spm} = \exp\big[-{\textstyle \frac i2}(\ga-\tfrac12)(e_\mp' \pp - e_\pm \pp')\big] \ .
\end{equation}
They can be written explicitly in terms of $x_\spm^\pm$ using \rf{xxq}.

As for the four phases $ \So{}_{_\pm} $, $\Sio{}_{_\pm}$, like their $q=0$ limits \ci{bssst} in \rf{ses},
they are not fixed by
the symmetry or the Yang-Baxter equation.
Observing that $(x^\pm_\spm)^* = x^\mp_\spm$, it can be seen that the S-matrix in the $q \neq 0$ case is QFT unitary,
while for braiding unitarity the phases should satisfy additional constraints, analogous to \eqref{pha2} in the $q=0$ case,
\begin{equation}\begin{split}
& \So{}_{_\pm}(x_\spm^+,x_\spm^-;\xpr_\spm^+,\xpr_\spm^-)\ \So{}_{_\pm}(\xpr_\spm^+,\xpr_\spm^-;x_\spm^+,x_\spm^-) = 1 \ ,
\\& \Sio{}_{_\pm}(x_\spm^+,x_\spm^-;\xpr_\smp^+,\xpr_\smp^-)\ \Sio{}_{_\mp}(\xpr_\smp^+,\xpr_\smp^-;x_\spm^+,x_\spm^-)
= \sqrt{\frac{x_\spm^+ \xpr_\smp^+}{x_\spm^-\xpr_\smp^-}} \, \frac{1-x_\spm^-\xpr_\smp^-}{1-x_\spm^+\xpr_\smp^+}\ .\la{pha3}
\end{split}\end{equation}
Furthermore, setting, as in eq.\eqref{form},
\begin{equation}\label{formq}
\a_\pm = e^{i\g\ov 2}\, \big(\frac{x^+_\spm}{x^-_\spm}\big)^\beta \ ,
\end{equation}
the crossing symmetry of the $q=0$ case, \eqref{cross1},\eqref{cross2}, also generalizes to $q \neq 0$
in a natural way with the ``crossed'' Zhukovsky variables \eqref{czv} given by
\begin{equation}
\bar x^\pm_\spm = \frac{1}{x^\pm_\smp} \ . \la{kr}
\end{equation}
Indeed, in view of \rf{444} the relations
\rf{kr} are equivalent to the expected relations for the ``crossed'' energy and momentum:
\begin{equation}
\bar e_\pm = - e_\mp \ , \qquad
\bar \pp = - \pp \ .
\end{equation}
Explicitly, so long as the four phases are related in the following way (analogous to \eqref{cross1})
\begin{equation}\begin{split}
\So^c{}_{_\pm}^{\vphantom{c}} = \Sio^{\vphantom{c}}{}_{_\pm}^{\vphantom{c}} \sqrt{\frac{x_\spm^-}{x_\spm^+}}\,
\frac{1-x_\spm^+\xpr_\smp^-}{1-x_\spm^-\xpr_\smp^-}\ , \qquad
\Sio^c{}_{_\pm}^{\vphantom{c}} = \So^{\vphantom{c}}{}_{_\mp}^{\vphantom{c}} \sqrt{\frac{x_\smp'^+}{x_\smp'^-}}\,
\frac{x_\smp^- - \xpr_\smp^-}{x_\smp^--\xpr_\smp^+}\ , \label{cross1a}
\end{split}\end{equation}
then
\begin{align}
L_{1_{\pm}}^c = & L_{2_{\pm}} ,
&
\Lambda_{1_{\pm}}^c= & \Lambda_{2_{\pm}} ,
&
L_{3_{\pm}}^c = & \Lambda_{6_\pm} ,
&
\Lambda_{3_\pm}^c = & L_{6_\pm} ,
&
L_{5_\pm}^c = & - i e^{i\g} \Lambda_{4_\pm} ,
&
\Lambda_{5_\pm}^c = & - i e^{-i\g} L_{4_\pm} ,\nonumber
\\
L_{2_\pm}^c = & L_{1_\mp} ,
&
\Lambda_{2_\pm}^c = & \Lambda_{1_\mp} ,
&
\Lambda_{6_\pm}^c = & L_{3_\mp} ,
&
L_{6_\pm}^c = & \Lambda_{3_\mp} ,
&
\Lambda_{4_\pm}^c = & i e^{-i\g} \Lambda_{5_\mp} ,
&
L_{4_\pm}^c = & i e^{i\g}L_{5_\mp} ,
\label{cross2q}
\end{align}
where $ \So^{c}{}_{_\pm}^{\vphantom{c}} = \So{}_{_\pm}(\bar x'^+_\spm,\bar x'^-_\spm;x^+_\spm,x^-_\spm) $
(likewise for
$L_{1,3,5_\pm},\Lambda_{1,3,5_\pm}$)
while $\Sio^{c}{}_{_\pm}^{\vphantom{c}} = \Sio{}_{_\pm}(\bar x'^+_\spm,\bar x'^-_\spm;x^+_\smp,x^-_\smp) $ (likewise for
$L_{2,4,6_\pm},\Lambda_{2,4,6_\pm}$).

\

It is then natural to conjecture that
the pattern of the generalization
to the $q \neq 0$ case described above may also apply to the phases, i.e.
to find their expressions in terms of the new Zhukovsky variables we just need to
replace $x^\pm \to x^\pm_\spm$ and $\xpr^\pm \to \xpr^\pm_\spm$ in the $q=0$ phases
as
\begin{equation}\label{phases}
\So{}_{_\pm} \stackrel{?}{=} \So(x^+_\spm,x^-_\spm;\xpr^+_\spm,\xpr^-_\spm)
\ , \qquad\ \ \ \ \
\Sio{}_{_\pm} \stackrel{?}{=} \Sio(x^+_\spm,x^-_\spm;\xpr^+_\smp,\xpr^-_\smp) \ .
\end{equation}
However, this prescription is ambiguous: since the dispersion relation is modified for $q\not=0$
(cf. \rf{yy} and \rf{4_7}), starting with
two expressions equal at $q=0$, using in one of them the $q=0$ dispersion relation and
then generalizing to $q\not=0$ as in \eqref{phases} we would find different
results. This suggests that the expressions
for the four undetermined phases should be given by some modification of
\rf{phases} that resolves this ambiguity.

In this paper we will present
such modification only
for the strong coupling limit of the phases.
In the $q=0$ case, at the leading order in $\hh\to \infty $
limit
the two phases $\So,\, \Sio$ were proposed \ci{bssst} to be equal to the AFS \ci{afs} phase (up to factors).
Explicitly, for $\So$ one has
\begin{align}\la{c}
&\So A\big|_{a=0} = \ \sqrt{\frac{x^+\xpr^-}{x^-\xpr^+}\ \frac{x^- - \xpr^+}{x^+ - \xpr^-}\ \frac{1-\frac{1}{x^-\xpr^+}}{1-\frac1{x^+\xpr^-}} }
\ \ \s_{_{\rm AFS}}^{-1}\ ,
\\
& \sigma_{_{\rm AFS}} ( x^+, x^-; x'^+, x'^-)= \ B\, e^{ i \hh\, \vartheta_0 } \ , \ \ \ \ \ \qquad \ \
B= { 1 - { 1 \ov x^- x'^+} \ov 1 - { 1 \ov x^+ x'^-} }
\ , \la{b}\\
\vartheta_0= & \ { 1 \ov 4} \big( x^+ + {1 \ov x^{+}} + x^- + {1 \ov x^{-}} - x'^+ - { 1 \ov x'^{+}} - x'^- - { 1 \ov x'^{-} } \big)
\log \Big[ { 1 - { 1 \ov x^+ x'^-} \ov 1 - { 1 \ov x^+ x'^+} } \, { 1 - { 1 \ov x^- x'^+}
\ov 1 - { 1 \ov x^- x'^-} }\Big]\ .
\la{bc}
\end{align}
Our proposal for $\So{}_{_\pm}$ for generic $q$ at strong coupling
is then given by the rule in eq.\eqref{phases} applied to \rf{c}
written in the form \eqref{b},\eqref{bc}, along with an additional simple modification --
the introduction of a factor of $\hq$ in front of $\vartheta_0$:
\begin{equation}
\sigma_{_{\rm AFS}} \ \to\ \sigma_{_{\rm AFS_q}} =B \, e^{i\hh\,\hq\,\vartheta_0}\ . \la{afs}
\end{equation}
Equivalently, one is to start from \rf{c} with $ \sigma_{_{\rm AFS}} $ given by \rf{afs}
and then apply the replacement in \rf{phases}.
Note that the introduction of the
extra
factor $\hq=\sqrt{1-q^2}$ in the exponent
(which may be motivated\,\foot{Starting with the dispersion relation \rf{4_7} we may formally introduce new variables
$y^\pm$ (we suppress the $_\pm$ subscripts, choosing, e.g., sign $+$ in \rf{4_7})

$
y^+ + {1\ov y^+} = \hat q(x^+ + {1\ov x^+}) - q(U - U^{-1}) = u + i\hh^{-1} , \ \ \ \
y^- + {1\ov y^- }= \hat q(x^- + {1\ov x^-}) - q(U^{-1} - U) = u - i \hh^{-1} , \ \ \ \ U = \sqrt{x^+\ov x^-} ,
$
such that $y^\pm = x^\pm$ when $q=0$. Then
the combination that should appear in front of log in $\vartheta_0$ in \rf{bc} is

$ y^+ + {1\ov y^+ } + y^- + {1\ov y^-} = \hat q(x^+ + {1\ov x^+} + x^- + {1\ov x^-})$.\fnsv
}
by the presence of $\hq$ in \rf{4_7}),
is still consistent with the QFT unitarity and braiding unitarity \eqref{pha3} of the resulting S-matrix.

Finding $\So{}_{_\pm}$ according to this prescription and
using the crossing relations \eqref{cross1a} to obtain the expressions for $\Sio{}_{_\pm}$\footnote{Note that because of the ambiguity discussed
beneath eq.\eqref{phases}, to find $\Sio{}_{_\pm}$ we cannot use the same procedure as used for $\So{}_{_\pm}$
starting with the $q=0$ expression given in \cite{bssst}.\fnsv}
we have checked explicitly that
in the BMN limit \eqref{bmn}, setting $\alpha_\pm = 1 + \mathcal{O}(\hh^{-1})$, we then
recover the tree-level string-theory S-matrix for the elementary massive excitations
(summarized in section \ref{sec1}) as found in \cite{htb} .

To compare to semiclassical string theory
one is to consider another strong-coupling limit -- the ``giant magnon'' limit:
\be \hh \to \infty\ , \ \ \ \ \ \ \pp={\rm fixed} \ , \ \ \ \ \ q={\rm fixed} \ . \la{gila} \ee
While the leading term \rf{gia} in the energy \rf{ds} does not depend on $q$ in this limit,
the Zhukovsky variables do. Expanding \rf{444} we find
\be
x_\spm^\pm = r_\spm \, e^{\pm { i \pp \ov 2} } \ , \ \ \ \ \ \ \ \
r_\spm = { 1 \ov \hq} \big[ 1 \pm q + { \hh^{-1} \ov 2 \sin { \pp \ov 2}} + O( \hh^{-2}) \big] = { 1 \ov \sqrt{ 1 \mp q} } + O(\hh^{-1})
\ .\label{24}
\ee
Taking this limit in the phase factors $\So{}_{_\pm}$, $\Sio{}_{_\pm}$
computed at $\hh\to \infty$ as explained below \rf{afs}, we find the following leading behaviour
\begin{equation}
\la{fa}
\So{}_{_\pm} \sim \Sio{}_{_\pm} \sim \ e^{- i \hh \, \hq \, \vartheta_0} \ , \ \ \ \ \ \ \ \ 
\vartheta_0 = \frac1\hq \, (\cos \frac{\pp}{2}-\cos \frac{\pp'}{2}) \, 
\log \frac{1 - \hq^2 \cos^2 \frac{\pp-\pp'}{4}}{1 - \hq^2 \cos^2 \frac{\pp+\pp'}{4}} \ + \mathcal O(\hh^{-1} )
\ ,
\end{equation}
where as always $\hq^2 = 1 - q^2$.
This expression is invariant under the crossing transformation $\pp \to - \pp'$, $\pp' \to \pp$ as expected.

For $q=0$ eq.\rf{fa} reduces to the familiar semiclassical ``giant-magnon'' limit \ci{hma} of the AFS phase.
For $q=1$ eq.\rf{fa} implies that the leading term in the phase is trivial. It would be interesting to
confirm this directly by considering ``giant magnon'' scattering in the WZW model.

\def \rmr {{\rm r}}
\def \ny {{\rm x}}
\def \nypr {{{\rm x}'}}
\def \Lt {\textrm{\small{L}}}
\def \Rt {\textrm{\small{R}}}
\subsection{$q=1$ case}\label{qeq1}

Let us now discuss explicitly the $q\to 1$ limit of the proposed exact S-matrix.
We first consider this limit in the dispersion relation \eqref{iq}.
Doing so we see that there are two different kinematic regions in \eqref{wzw}:
\begin{equation}\begin{split}\label{reg}
(i) \ \ 1\pm 2 \hh\,\sin \frac\pp 2 > 0 \quad & \Rightarrow \quad e_\pm = 1 \pm 2\hh \sin\frac\pp2 \ ,
\\
(ii) \ \ 1\pm 2 \hh\,\sin \frac\pp 2 < 0 \quad & \Rightarrow \quad e_\pm = - 1 \mp 2\hh \sin\frac\pp2
\ .
\end{split}\end{equation}
These two regions are a ``discrete'' generalization of the left- and right-moving dispersion relations in a
massless 2-d relativistic theory.
From \eqref{444} we see that in the $q \to 1$ limit the behaviour of $x^\pm_\spm$ is different in the two cases \eqref{reg}:
\begin{equation}
(i) \ \ x_\spm^\pm \sim \hq^{-1} \ , \qquad\qquad
(ii) \ \ x_\spm^\pm \sim \hq \ .
\end{equation}
Defining the following rescaled variables
\begin{equation} \la{qaq} \ \ \ \ \ \ \
(i) \ \ \ny^\pm_\spm = \hq\, x^\pm_\spm \ ,\ \ \ \ \ \qquad (ii) \ \ \ny^\pm_\spm = \hq^{-1} \, x^\pm_\spm\ ,
\end{equation}
which are finite in the $q \to 1$ limit, and using them in the dispersion relation \eqref{4_7} written in terms
of the Zhukovsky variables, we find the following $q \to 1$ limits of \eqref{4_7}
\begin{equation}\begin{split}\label{5_7}
(i) \ \ (\ny_\spm^+ - \ny_\spm^- )\Big ( 1 \mp { 2 \ov \sqrt{ \ny_\spm^+\ny_\spm^-}}\Big) =\frac{2i}{\hh}\ ,\qquad
(ii) \ \ \Big(\frac1{\ny_\spm^+} - \frac1{\ny_\spm^-}\Big)(1 \pm 2 \sqrt{\ny_\spm^+ \ny_\spm^-}) = \frac{2i}{\hh}\ .
\end{split}\end{equation}
The $q \to 1$ limit of \eqref{444} is then
\begin{equation}\begin{split}
\ny_\spm^\pm = {\rmr}_\spm \, e^{\pm \frac{i\pp}{2}} \ , \ \ \ \ \
(i) \ \ {\rmr}_\spm = \frac{
1 \pm 2\hh \,\sin \frac \pp2}{
\hh\,\sin\frac \pp2} \ , \ \ \ \ \
(ii) \ \ {\rmr}_\spm = - 
\frac{
\hh\,\sin\frac\pp2}{
1 
\pm 2\hh \,\sin \frac \pp2}
\ ,
\label{445}
\end{split}\end{equation}
where we have used the expressions for the energy in terms of momentum given in eq.\eqref{reg}. 
Substituting \eqref{445} into \eqref{5_7} we recover the $q \to 1$ limit of the dispersion relation \eqref{reg} in the two
regions as expected.

Taking the $q \to 1$ limit in the S-matrix, using the rescaled variables \eqref{qaq}, there are four possibilities corresponding to the unprimed and primed momenta either being in the region $(i)$ or $(ii)$. Here we give two examples of the limit:
the first is when both excitations have momentum in region $(i)$ and the second is
when the unprimed excitation has momentum in region $(i)$ and the primed excitation has momentum in region $(ii)$.
Assuming that $\a$ has the form given in eq.\eqref{formq}, in the first
case the $q \to 1$ limit of the parametrizing functions $L_{1,3,5_\pm},\Lambda_{1,3,5_\pm}$
take the same form as in \eqref{sqs}, just with $x^\pm_\spm \to \ny^\pm_\spm$, while for the remaining functions we find
\begin{align}
L_{6_{\pm}} = \ & \Sio{}_{_\pm} \ , & \no
L_{2_\pm} = \ & {\textstyle \Sio{}_{_\pm} \sqrt{\frac{\ny_\spm^+}{\ny_\spm^-}} } \ , &
L_{4_{\pm}} = \ & 0 \ , \\
\Lambda_{6_{\pm}} = \ &{ \textstyle \Sio{}_{_\pm} \sqrt{\frac{\ny_\spm^+\nypr_\smp^+}{\ny_\spm^-\nypr_\smp^-}} } \ , &
\Lambda_{2_{\pm}} = \ & {\textstyle \Sio{}_{_\pm} \sqrt{\frac{\nypr_\smp^+}{\nypr_\smp^-}} } \ , &
\Lambda_{4_{\pm}} = \ & 0 \ .\label{sqsq2}
\end{align}
In the second case the parametrizing functions $L_{2,3,4_\pm}$, $\Lambda_{2,3,4_\pm}$
take the same form as in \eqref{sqs}, again with $x^\pm_\spm \to \ny^\pm_\spm$, while the remaining functions are
\begin{align}\no
L_{1_\pm} = \ & \So{}_{_\pm} \ , &
L_{3_{\pm}} = \ & {\textstyle \So{}_{_\pm} \sqrt{\frac{\ny_\spm^+}{\ny_\spm^-}} } \ , &
L_{5_{\pm}} = \ & 0 \ ,
\\
\Lambda_{1_{\pm}} = \ & {\textstyle \So{}_{_\pm} \sqrt{\frac{\ny_\spm^+\nypr_\spm^+}{\ny_\spm^-\nypr_\spm^-}} } \ , &
\Lambda_{3_{\pm}} = \ & {\textstyle \So{}_{_\pm} \sqrt{\frac{\nypr_\spm^+}{\nypr_\spm^-}} } \ , &
\Lambda_{5_{\pm}} = \ & 0 \ . \label{sqsq3}
\end{align}
The corresponding Bethe ansatz for the spectrum should then
have a substantial simplification in this limit.
In particular, taking the observation below eq.\rf{fa} as a hint, one may
expect that the exact phases should trivialize in the $q=1$ limit.
This would be in line with the expected simplification
of the spectrum in the case of the $AdS_3 \times S^3$ string theory with NSNS flux
which is described by the WZW theory (here in a light-cone type gauge).\foot{There is
an interesting open question
about the possible relation between this exact S-matrix appearing in the $q\to 1$ limit
for scattering of ``solitonic'' states with the dispersion relation \rf{wzw}
and the massless S-matrices for scattering of elementary excitations in the $k=1$ \ci{zz} and $k > 1$ \ci{ber}
$SU(2)$ WZW model.\fnsv}

\section{Concluding remarks}\label{secconc}

The exact dispersion relation and the S-matrix presented above is a starting point
for the construction of the corresponding Bethe ansatz for the string spectrum in the general $q\not=0$ case.
The full S-matrix is a product \ci{htb,bssst} of two copies of the
``elementary'' S-matrices \rf{bct} with the coefficient
functions given in \rf{sqs}. Remarkably, these are exactly the same as in the $q=0$ case \rf{ses} \ci{bssst},
just with $\spm$ subscripts added. The details are then encoded in the generalization of the dispersion relation
to the $q\not=0$ case according to \rf{iq},\rf{4_7}.

This suggests that the corresponding Bethe ansatz that corresponds to this scattering matrix
should have essentially the same structure as found in the $q=0$ case in \ci{bssst}.
Once again, this is largely due to the symmetry algebra being the same for any value of $q$.\foot{In this
sense, the case of $AdS_3 \times S^3 \times S^3\times S^1$ theory appears to be a more complicated
1-parameter generalization as there the symmetry algebra depends on the deformation parameter $\a$ \ci{bsz,bor1}.\fnsv}
The same should apply to the construction of the corresponding Y-system and TBA equations.

One outstanding open problem (already for $q=0$)
is to find the exact expressions for the four dynamical phases
$ \So{}_{_\pm}, \Sio{}_{_\pm} $ in the S-matrix \rf{sqs}. As discussed in section \ref{qneq1} below eq.\rf{phases},
their generalization from $q=0$ to $q\not=0$ may not be as straightforward as for the S-matrix coefficients in
\rf{ses},\rf{sqs}.

There are a number of additional investigations that are required
to check the formal construction of this paper against perturbative string theory.
First, one should match the
semiclassical strong-coupling limit of the BA equations corresponding to
the S-matrix with the phases given by \eqref{afs} with the finite-gap
equations for the corresponding classical string sigma model, generalizing the discussion
in the $q=0$ case in \ci{bssst}.
It would be interesting also to derive the semiclassical phase \rf{fa}
from ``giant magnon'' scattering considerations as in \ci{hma}.

It is also important to find the string one-loop (and possibly two-loop) corrections to the tree-level BMN
S-matrix to determine the corresponding subleading terms in the four phases.
Such computations appear to be feasible
using unitarity techniques recently described in \ci{bhf,roi}.
One simplifying option is to consider the analog of the near-flat limit as was done
in the $q=0$ case in \ci{wu}.
It should be possible also to find the one-loop corrections
to phases by studying the leading quantum corrections
near semiclassical solutions like the ``giant magnon'' and
spinning string, generalizing the corresponding investigations
\ci{david,bec} in the $q=0$ case.

It would be interesting also to construct an exact 3-parameter $(\hh,q, \rm k)$
S-matrix that interpolates (as in the $AdS_5\times S^5$ case \ci{hhm})
between the exact superstring S-matrix parametrized by $(\hh,q)$ and
the exact relativistic S-matrix of the corresponding Pohlmeyer-reduced theory \ci{htb,ht2,hhm1}
parametrized by $(q,\rm k)$.\foot{Here $\rm k$ stands for the coefficient in front of the action
of the Pohlmeyer-reduced theory.\fnsv}
As was shown in \ci{htb}, the Pohlmeyer-reduced theory in the $q\not=0$ case depends on $q$ only through
the mass scale $\bar \mu = \hq \mu$, i.e. its relativistic S-matrix is actually independent of $q$.
The interpolating S-matrix should have a non-trivial dependence on $q$
to recover the superstring S-matrix in the appropriate limit.

\

Finally, let us mention that it would be interesting to generalize the construction of this paper
to the case of superstring theory on $AdS_3 \times S^3 \times S^3 \times S^1$ supported by a mixture
of RR and NSNS flux parametrized by the two parameters $ \a$ and $q$.
The $AdS_3 \times S^3 \times S^3 \times S^1$ supported just by RR flux ($q=0$) depends on
$\a\in (0,1) $ \cite{bsz,b1,bor1}\,\footnote{The radii of the two spheres are given by $R_1^2 = \alpha^{-1}, \ R_2^2 = ( 1 - \alpha)^{-1}$.\fnsv}
so one may compare this one-parameter generalization of $AdS_3 \times S^3 \times T^4$
with the one (with NSNS flux parameter $q$) we discussed in this paper.
One obvious difference is that the two theories have different symmetry algebras: $\mathfrak{d}(2,1;\a)^2$ and $\mathfrak{psu}(1,1|2)^2$.
In both cases the dispersion relation is modified according to the construction in footnote \ref{footdis}.
However, in contrast to the $AdS_3 \times S^3 \times S^3 \times S^1$ case for which $M_+$ and $M_-$ remain equal, the introduction of $q$
lifts the ``degeneracy'' between the representation parameters with $+$ and $-$ subscripts, i.e. $M_+ \neq M_-$.\footnote{This
is a consequence of the fact that parity symmetry is broken with the introduction of the NSNS flux. However, charge conjugation composed
with parity is still a symmetry.\fnsv}
Furthermore, in the $AdS_3 \times S^3 \times S^3 \times S^1$ case $M=M_\pm$ takes the constant value $\a$ or $1-\a$, while
here (see \rf{Functions}) it has a dependence on the momentum and thus may have a non-trivial effect on the analytic structure.
By combining the features of the two ($q=0$ and $\a=0$) constructions, it should be relatively simple to suggest a proposal for
the exact massive S-matrix for the superstring theory on
$AdS_3 \times S^3 \times S^3 \times S^1$ supported by a mixture
of RR and NSNS flux.

\

\section*{Acknowledgments}

We would like to thank A. Babichenko, M. Pawellek, R. Roiban and L. Wulff for related discussions. We also thank
A. Babichenko, R. Roiban and B. Stefanski for useful comments on the draft.
BH is supported by the Emmy Noether Programme ``Gauge fields from Strings'' funded by the German Research Foundation (DFG).
The work of AAT is supported by the ERC Advanced grant No.290456 and also by the STFC grant ST/J000353/1.



\end{document}